\documentclass{sig-alternate}
\usepackage{balance,array,microtype,pifont,enumitem}
\usepackage{floatrow}
\usepackage[font=bf]{caption}
\usepackage{subfig}

\usepackage{color, times, mathptmx, amsmath, amssymb, amsthm, verbatim, 
  mdwlist, graphicx, footnote, multirow, xspace, tikz, colortbl,dcolumn, mathtools}

\usepackage[noadjust]{cite}

\DeclareMathAlphabet{\mathcal}{OMS}{cmsy}{m}{n}
\let\footnotesize\small

\usepackage{algorithm}
\usepackage[noend]{algpseudocode}

\algrenewcommand{\alglinenumber}[1]{\small#1:}

\theoremstyle{definition}
\usepackage[hidelinks]{hyperref}
\usepackage{breakurl}

\newcommand{\macrobase}{MacroBase\xspace}
\newcommand{\cmt}{CMT\xspace}
\newcommand{\mdp}{MDP\xspace}
\newcommand{\cambridgemobiletelematics}{Cambridge Mobile Telematics\xspace}

\newcommand{\cmtd}{CMT\xspace}
\newcommand{\campaigns}{{Campaign}\xspace}
\newcommand{\disburse}{{Disburse}\xspace}
\newcommand{\accidents}{{Accidents}\xspace}
\newcommand{\telecom}{{Telecom}\xspace}

\newcommand{\liquor}{{Liquor}\xspace}

\newcommand{\minihead}[1]{{\vspace{.45em}\noindent\textbf{#1.} }}

\newif\ifextended
\extendedtrue

\usepackage{etoolbox}
\newcommand{\zerodisplayskips}{%
  \setlength{\abovedisplayskip}{4pt}
  \setlength{\belowdisplayskip}{4pt}
  \setlength{\abovedisplayshortskip}{4pt}
  \setlength{\belowdisplayshortskip}{4pt}}
\appto{\normalsize}{\zerodisplayskips}
\appto{\small}{\zerodisplayskips}
\appto{\footnotesize}{\zerodisplayskips}

\usepackage{inconsolata}

\begin{document}
%

\CopyrightYear{2017} 
\isbn{978-1-4503-4197-4/17/05}
\doi{http://dx.doi.org/10.1145/3035918.3035928}


\clubpenalty=10000 
\widowpenalty = 10000

\title{\macrobase: Prioritizing Attention in Fast Data}

{\author{Peter Bailis, Edward Gan, Samuel Madden$^\dagger$, Deepak Narayanan,
    Kexin Rong, Sahaana Suri \\[1.5mm]
\affaddr{Stanford InfoLab and $^\dagger$MIT CSAIL}}
\maketitle


\begin{abstract}
As data volumes continue to rise, manual inspection is becoming increasingly untenable.
In response, we present \macrobase, a data analytics engine that prioritizes end-user attention in high-volume \textit{fast data} streams.
\macrobase enables efficient, accurate, and modular analyses that highlight and aggregate important and unusual behavior, acting as a search engine for fast data. 
\macrobase is able to deliver order-of-magnitude speedups over alternatives by optimizing the combination of explanation and classification tasks and by leveraging a new reservoir sampler and heavy-hitters sketch specialized for fast data streams.
As a result, \macrobase delivers accurate results at speeds of up to 2M events per second per query on a single core. The system has delivered meaningful results in production, including at a telematics company monitoring hundreds of thousands of vehicles.
 \end{abstract}

\section{Introduction}
\label{sec:intro}

Data volumes are quickly outpacing human abilities to process them.
Today, Twitter, LinkedIn, and Facebook each record over 12M events per second~\cite{twitter-volume,linkedin-volume,fb-volume}. 
These volumes are growing and are becoming more common: machine-generated data sources such as sensors, processes, and automated systems are projected to increase data volumes by 40\% each year~\cite{emc}. 
However, human attention remains limited; it is becoming increasingly impossible to rely on manual inspection and analysis of these large data volumes. They are simply too large.
Due to this combination of immense data volumes and limited human attention, today's best-of-class application operators anecdotally report accessing less than 6\% of data they collect~\cite{mb-cidr}, primarily in reactive root-cause analyses.

While humans cannot manually inspect these \textit{fast data} streams,
machines can~\cite{mb-cidr}.  Machines can filter, highlight, and
aggregate fast data, winnowing and summarizing data before it reaches
a user.  As each result shown to the end-user consumes their
attention~\cite{simon}, we can help prioritize this attention by
leveraging computational resources to maximize the utility of each
result shown.  That is, fast data necessitates a search engine to help
identify the most relevant data and trends (and to allow non-expert
users to issue queries). The increased availability of elastic
computation as well as advances in machine learning and statistics
suggest that the construction of such an engine is possible.


However, the design and implementation of this infrastructure is
challenging; current analytics deployments are a far cry from this
potential. Today, application developers and analysts can employ a
range of scalable dataflow processing engines to compute over fast
data (over 20 in the Apache Software Foundation alone). However, these
engines leave the actual implementation of scalable analysis operators
that prioritize attention (e.g., highlighting, grouping, and
contextualizing important behaviors within fast data) up to the
application developer. This development is hard: fast data analyses
must $i.)$ determine the few results to return to end users (to avoid
overwhelming their attention) while $ii.)$ executing quickly to keep
up with immense data volumes and $iii.)$ adapting to changes within
the data stream itself. Thus, designing and implementing these
analytics operators requires a combination of domain expertise,
statistics and machine learning, and dataflow processing. This
combination is rare. Instead, today's high-end industrial deployments
overwhelmingly rely on a combination of static rules and thresholds
that analysts report are computationally efficient but brittle and
error-prone; manual analysis is typically limited to reactive,
post-hoc error diagnosis that can take hours to days.

To bridge this gap between the availability of low-level dataflow
processing engines and the need for efficient, accurate analytics
engines that prioritize attention in fast data, we have begun the
development of MacroBase, a fast data analysis system. The core
concept behind MacroBase is simple: to prioritize attention, an
analytics engine should provide analytics operators that automatically
classify and explain fast data volumes to users.
MacroBase executes extensible streaming dataflow pipelines that
contain operators for both \textit{classifying} individual data points
and \textit{explaining} groups of points by aggregating them and
highlighting commonalities of interest. Combined, these operators
ensure that a few returned results capture the most important
properties of data. Much as in conventional relational analytics, when
designed for reuse and composition, a small core set of efficient fast
data operators allows portability across application domains.

The resulting research challenge is to determine this efficient,
accurate, and modular set of core classification and explanation
operators for prioritizing attention in fast data. The statistics and
machine learning literature is replete with candidate algorithms, but
it is unclear which can execute online at fast data volumes, and, more
importantly, how these operators can be composed in an end-to-end
system.  Thus, in this paper, we both introduce the core MacroBase
architecture---which combines domain-specific feature extraction with
streaming classification and explanation operators---and present the
design and implemention of MacroBase's default streaming
classification and explanation operators. In the absence of labeled
training data, MacroBase executes operators for unsupervised, density-based
classification that highlight points lying far from the overall
population according to user-specified \textit{metrics} of interest
(e.g., power drain). MacroBase subsequently executes sketch-based
explanation operators, which highlight correlations that most
differentiate outlying data points according to their
\textit{attributes} (e.g., firmware version, device ID).

Users of the open source MacroBase
prototype\footnote{https://github.com/stanford-futuredata/macrobase}
have utilized MacroBase's classification and explanation operators to
find unusual and previously unknown behaviors in fast data from mobile
devices, datacenter telemetry, automotives, and manufacturing
processes, such as in the following example.

\vspace{.5em}\noindent\textit{\textsc{Example.} A mobile application
  manufacturer issues a \macrobase query to monitor power drain
  readings (i.e., \emph{metrics}) across devices and application
  versions (i.e., \emph{attributes}). \macrobase's default operator
  pipeline reports that devices of type B264 running application
  version 2.26.3 are sixty times more likely to experience abnormally
  high power drain than the rest of the stream, indicating a potential
  problem with the interaction between devices of type B264 and
  application version 2.26.3.}\vspace{.5em}

\noindent Beyond this basic default functionality, MacroBase allows users to tune
their queries by $i.)$ adding domain-specific feature transformations
(e.g., time-series operations such as Fourier transform and
autocorrelation) to their pipelines---without modifying the rest
of the pipeline, $ii.)$ providing supervised classification rules (or
labels) to complement or replace unsupervised classifiers and
$iii.)$ authoring custom streaming transformation, classification, and
explanation operators, whose interoperability is enforced by
\macrobase's type system and can be combined with
relational operators.

\vspace{.5em}\noindent\textit{\textsc{Example.}  The mobile
  application developer also wishes to find time-varying power spikes
  within the stream, so she reconfigures her pipeline by adding a
  time-series feature transformation to identify time periods with
  abnormal time-varying frequencies. She later adds a
  manual rule to capture all readings with power drain greater than
  $100$W and a custom time-series explanation operator~\cite{ts-vis}---all without modifying the remainder of the operator
  pipeline.}\vspace{.5em}

\noindent Developing these operators necessitated several algorithmic advances,
which we address as core research challenges in this paper:

To provide responsive analyses over dynamic data sources, \macrobase's
default operators are designed to adapt to shifts in data. \macrobase
leverages a novel stream sampler, called the \textit{Adaptable Damped
  Reservoir} (ADR), which performs sampling over arbitrarily-sized,
exponentially damped windows. \macrobase uses the ADR to incrementally
train unsupervised classifiers based on statistical density estimation
that can reliably identify typical behavioral modes despite large
numbers of extreme data points~\cite{robust-huber}. \macrobase also
adopts exponentially weighted sketching and streaming data
structures~\cite{cormode-decay,cps-tree} to track correlations between
attribute-value pairs, improving responsiveness and accuracy in explanation.


To provide interpretable explanations of often relatively rare
behaviors in streams, \macrobase adopts a metric from statistical
epidemiology called the \textit{relative risk ratio} that describes
the relative occurrence of key attributes (e.g., age, sex) among
infected and healthy populations. In computing this statistic,
\macrobase employs two new optimizations. First, \macrobase exploits
the cardinality imbalance between classified points to accelerate
explanation generation, an optimization enabled by the combination of
detection and explanation. Instead of inspecting ``outliers'' and
``inliers'' separately, \macrobase first examines the small set of
outliers, then aggressively prunes its search over the much larger set
of inliers. Second, \macrobase exploits the fact that many fast data
streams contain repeated measurements from devices with similar
attributes (e.g., firmware version) during risk ratio computation,
reducing data structure maintenance overhead via a new counting
sketch, the \textit{Amortized Maintenance Counter} (AMC). These
optimizations improve performance while highlighting the often small
subset of attributes that matter most.

We report on early production experiences and quantitatively evaluate
\macrobase's performance and accuracy on both production telematics
data as well as a range of publicly available real-world
datasets. \macrobase's optimized operators exhibit order-of-magnitude
performance improvements over existing operators at rates of up to 2M
events per second per query while delivering accurate results in
controlled studies using both synthetic and real-world data. As we
discuss, this ability to quickly process large data volumes can also
improve result quality: large numbers of samples combat statistical
bias due to the multiple testing problem~\cite{bonferroni}, thereby
improving result significance. We also demonstrate \macrobase's
extensibility via case studies in mobile telematics, electricity
metering, and video-based surveillance, and via integration with several
existing analytics frameworks.

We make the following contributions in this paper:
\begin{itemize}[itemsep=.1em,parsep=.4em,topsep=.5em]
\item \macrobase, an analytics engine and architecture for analyzing
  fast data streams that is the first to combine streaming outlier
  detection and streaming data explanation.

\item The Adaptable Damped Reservoir, the first exponentially damped
  reservoir sample to operate over arbitrary windows, which \macrobase
  leverages in online classifier training.

\item An optimization for improving the efficiency of combined
  detection and explanation by exploiting cardinality imbalance
  between classes in streams.

\item The Amortized Maintenance Counter, a new heavy-hitters sketch
  that allows fast updates by amortizing sketch pruning across multiple
  observations of the same item.

\end{itemize}

The remainder of this paper proceeds as
follows. Section~\ref{sec:background} describes our target environment
by presenting motivating use cases. Section~\ref{sec:architecture}
presents the \macrobase's interaction model and primary default analysis pipeline
(which we denote \mdp). Section~\ref{sec:detection} describes \macrobase's default
streaming classification operators and presents the ADR
sampler. Section~\ref{sec:summarization} describes \macrobase's
default streaming explanation operator, including its
cardinality-aware optimization and the AMC sketch. We experimentally
evaluate \macrobase's accuracy and runtime, report on experiences in
production, and demonstrate extensibility via case studies in
Section~\ref{sec:evaluation}. Section~\ref{sec:relatedwork} discusses
related work, and Section~\ref{sec:conclusion} concludes.

\section{Target Environment}
\label{sec:background}

\macrobase provides application writers and system analysts an
end-to-end analytics engine capable of classifying data within
high-volume streams while highlighting important properties of the
data within each class.  As examples of the types of workloads we seek
to support, we draw on three motivating use cases from industry.

\minihead{Mobile applications} \cambridgemobiletelematics (\cmt) is a five-year-old telematics company whose mission is to make roads safer by making drivers more aware of their driving habits. \cmt provides drivers with a smartphone application and mobile sensor for their vehicles, and collects and analyzes data from many hundreds of thousands of vehicles at rates of tens of Hz. \cmt uses this data to provide users with feedback about their driving. 

\cmt's engineers report that monitoring their application has proven especially challenging. \cmt's operators, who include database and systems research veterans, report difficulty in answering several questions: is the \cmt application behaving as expected?  Are all users able to upload and view their trips?  Are sensors operating at a sufficiently granular rate and in a power-efficient manner? The most severe problems in the \cmt application are caught by quality assurance and customer service, but many behaviors are more pernicious.  For example, Apple iOS 9.0 beta 1 introduced a buggy Bluetooth stack that prevented iOS devices from connecting to \cmt's sensors. Few devices ran these versions, so the overall failure rate was low; as a result, \cmt's data volume and heterogeneous install base (which includes the 24K distinct device types in the Android ecosystem) obscured a potentially serious widespread issue in later releases of the application. Given low storage costs, \cmt records all of the data required to perform analytic monitoring to detect such behaviors, yet \cmt's engineers report they have lacked a solution for doing so in a timely and efficient manner.

In this paper, we report on our experiences deploying \macrobase at \cmt, where the system has highlighted interesting behaviors such as those above, in production.

\minihead{Datacenter operation} Datacenter and server operation represents one of the highest-volume data sources today. In addition to the billion-plus events per minute volumes reported at Twitter and LinkedIn, engineers reported a similar need to quickly identify misbehaving servers, applications, and virtual machines.

For example, Amazon AWS recently suffered a failure in its DynamoDB service, resulting in outages at sites including Netflix and Reddit. The Amazon engineers reported that ``after we addressed the key issue...we were left with a low overall error rate, hovering between 0.15-0.25\%. We knew there would be some cleanup to do after the event,'' and therefore the engineers deferred maintenance. However, the engineers ``did not realize soon enough that this low overall error rate was giving some customers disproportionately high error rates'' due to a misbehaving server partition~\cite{amazon-postmortem}.

This public postmortem is representative of many scenarios described by system operators in interviews. At a major social network, engineers reported that the challenge of identifying transient slowdowns and failures across hosts and containers is exacerbated by the heterogeneity of workload tasks. Failure postmortems can take hours to days, and, due to the labor-intensive nature of manual analysis, engineers report an inability to efficiently and reliably identify slowdowns, leading to suspected inefficiency.

Unlike the \cmt use case, we do not directly present results over production data from these scenarios. However, datacenter telemetry is an area of ongoing activity within the \macrobase project.

\minihead{Industrial monitoring} Increased sensor availability has spurred interest in and collection of fast data in industrial deployments. While many industrial systems already rely on legacy analytics systems, several industrial application operators we encountered reported a desire for analytics and alerting that can adapt to new sensors and changing conditions. These industrial scenarios can have important consequences. For example, an explosion and fire in July 2010 killed two workers at Horsehead Holding Corp.'s Monaca, PA, zinc manufacturing plant. The US Chemical Safety board's postmortem revealed that ``the high rate-of-change alarm warned that the [plant] was in imminent danger 10 minutes before it exploded, but there appears to have been no specific alarm to draw attention of the operator to the subtle but dangerous temperature changes that were taking place much (i.e. hours) earlier.'' The auditor noted that ``it should be possible to design a more modern control system that could draw attention to trends that are potentially hazardous''~\cite{horsehead}.

In this paper, we illustrate the potential to draw attention to unusual behaviors within electrical utilities.

\section{\macrobase Architecture and APIs}
\label{sec:architecture}

\begin{table}
\small
\center
\begin{tabular}{|l|l |}
\hline

\multicolumn{2}{|l|}{\textsc{Data Types}} \\
  
\multicolumn{2}{|l|}{$Point$ := (\texttt{array<double>} metrics,
  \texttt{array<varchar>} attributes)} \\ 

\multicolumn{2}{|l|}{$Explanation$ := (\texttt{array<varchar>} attributes,
  $stats$ statistics)
  } \\\hline \hline

\multicolumn{2}{|l|}{\textsc{Operator Interface}} \\

\multicolumn{1}{|l}{\textit{Operator}} & \multicolumn{1}{l|}{\textit{Type Signature}} \\ \hline

Ingestor & external data source(s) $\rightarrow$ \texttt{stream<$Point$>}\\

Transformer & stream<$Point$> $\rightarrow$ \texttt{stream<$Point$>}\\

Classifier & stream<$Point$> $\rightarrow$ \texttt{stream<(label, $Point$)>}\\

Explainer & \texttt{stream<(label, $Point)$>} $\rightarrow$

\texttt{stream<$Explanation$>}\\\hline\hline

Pipeline & Ingestor $\rightarrow$ \texttt{stream<$Explanation$>} \\\hline

\end{tabular}
\caption{\macrobase's core data and operator types. Each
  operator implements a strongly typed, stream-oriented
  dataflow interface specific to a given pipeline
  stage. A pipeline can utilize multiple operators
  of each type via transformations, such as group-by
  and one-to-many stream replication, as long as the pipeline
  ultimately returns a single stream of explanations. 
  }
\label{table:api}
\end{table}

As a fast data analytics engine, \macrobase filters and aggregates
large, high-volume streams of potentially heterogeous data. As a
result, \macrobase's architecture is designed for high-performance
execution as well as flexible operation across domains using an array
of classification and explanation operators. In this section, we
describe \macrobase's query processing architecture, approach to
extensibility, and interaction modes.

\subsection{Core Concepts}

To prioritize attention, MacroBase executes streaming analytics
operators that help filter and aggregate the stream. To do so, it
combines two classes of operators:

\minihead{Classification} Classification operators examine individual
data points and label them according to user-specified classes. For
example, MacroBase can classify an input stream of power drain
readings into two classes: points representing statistically
``normal'' readings and abnormal ``outlying'' readings. \vspace{.5em}

At scale, surfacing even a handful of raw data points per second can
overwhelm end users, especially if each data point contains
multi-dimensional and/or categorical information. As a result,
MacroBase employs a second type of operator:\vspace{.5em}

\minihead{Explanation} Explanation operators group and aggregate
multiple data points. For example, MacroBase can describe
commonalities among points in a class, as well as differences between
classes. Each result returned by an explanation operator can represent
many individual classification outputs, further prioritizing
attention. \vspace{.5em}

As we discuss in Section~\ref{sec:relatedwork}, classification and
explanation are core topics in several communities including
statistics and machine learning. Our goal in MacroBase is to develop
core operators for each task that are able to execute quickly over
streaming data that may change over time and can be composed as part
of end-to-end pipelines. Conventional relational analytics have a
well-defined set of composable, reusable operators; despite pressing
application demands at scale, the same cannot be said of
classification and explanation today. Identifying these operators and
combining them with appropriate domain-specific {feature extraction}
operators enables reuse beyond one-off, ad-hoc analyses.

Thematically, our focus is on developing operators that deliver more
information using less output. This score-and-aggregate strategy is
reminiscent of many data-intensive domains, including search. However,
as we show, adapting these operators for use in efficient, extensible
fast data pipelines requires design modifications and even enables new
optimizations. When employed in a system designed for extensibility, a
small number of optimized, composable operators can execute across
domains.

\subsection{System Architecture}

\minihead{Query pipelines} \macrobase executes pipelines of
specialized dataflow operators over input data streams. Each
\macrobase \textit{query} specifies a set of input data sources as
well as a logical query plan, or \textit{pipeline} of streaming
operators, that describes the analysis.

\macrobase's pipeline architecture is guided by two principles. First, all
operators operate over streams. Batch execution is supported by
streaming over stored data. Second, \macrobase uses the compiler's
type system to enforce interoperability. Each operator
must implement one of several type signatures (shown in
Table~\ref{table:api}). In turn, the compiler enforces that all
pipelines composed of these operators will adhere to the common
structure we describe below.

This \textit{architecture via typing} strikes a balance between the
elegance of more declarative but often less flexible interfaces and
the expressiveness of more imperative but often less composable
interfaces. More specifically, this use of the type system facilitates
three important kinds of interoperability. First, users can substitute
streaming detection and explanation operators without concern for
their interoperability. Early versions of the \macrobase prototype
that lacked this modularity were hard to adapt. Second,
users can write a range of domain-specific feature transformation
operators to perform advanced processing
(e.g., time-series operations) without requiring
expertise in classification or explanation. Third, \macrobase's
operators preserve compatibility with dataflow operators found in
traditional stream processing engines. For example, a \macrobase
pipeline can contain standard selection, project, join, windowing,
aggregation, and group-by operators.

\begin{figure*}
\includegraphics[width=\textwidth]{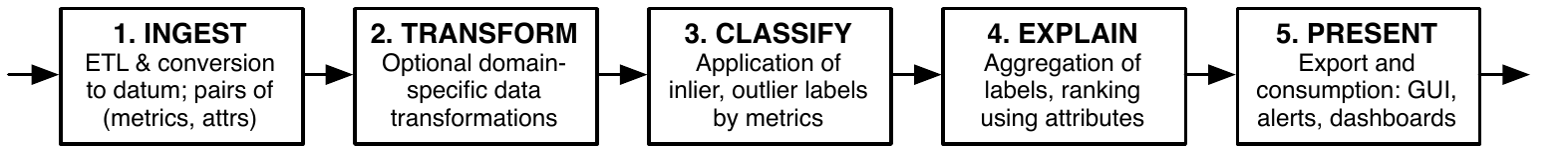}\vspace{-0.5em}
\caption{\macrobase's default analytics pipeline: \macrobase ingests
  streaming data as a series of points, which are scored and
  classified, aggregated by an explanation operator, then ranked and
  presented to end users. }
\label{fig:hilevel}
\end{figure*}

A \macrobase pipeline is
structured as follows:

\minihead{1.) Ingestion} \macrobase ingests data streams for analysis
from a number of external data sources. For example, \macrobase's JDBC
interface allows users to specify columns of interest from a base view
defined by a SQL query. \macrobase subsequently reads the result-set
from the JDBC connector, and constructs the set of data points to
process, with one point per row in the view. \macrobase currently
requires that any necessary stream ordering and joins be performed by
this initial ingestion operator.

Each data point contains a set of \textit{metrics}, corresponding to
key measurements (e.g., trip time, battery drain), and
\textit{attributes}, corresponding to associated metadata
(e.g., user ID and device ID). \macrobase uses metrics to detect
abnormal or unusual events, and attributes to explain behaviors. In
this paper, we consider real-valued metrics and categorical
attributes.\footnote{We discretize continuous attributes
  (e.g., see ~\cite{dbsherlock}) and provide two examples of discretization
  in Section~\ref{sec:extensibility}.}
 
As an example, to detect the OS version problem at \cmt, trip times could be used as a metric, and device and OS
type as attributes. To detect the outages at DynamoDB, error rates could be used as a metric, and server or IP address
as an attribute. To detect the Horsehead pressure losses, pressure gauge readings could be used as metrics and their
locations as attributes, as part of an autocorrelation-enabled
time-series pipeline (Section~\ref{sec:extensibility}). Today,
selecting attributes, metrics, and a pipeline is a user-initiated
process; ongoing extensions (Section~\ref{sec:conclusion}) seek to
automate this.

\minihead{2.) Feature Transformation} Following ingestion, \macrobase
executes an optional series of domain-specific data transformations
over the stream, which could include time-series specific operations
(e.g., windowing, seasonality removal, autocorrelation, frequency
analysis), statistical operations (e.g., normalization, dimensionality
reduction), and datatype specific operations (e.g., hue extraction for
images, optical flow for video).  For example, in
Section~\ref{sec:casestudy}, execute a pipeline containing a grouped
Fourier transform operator that aggregates the stream into hour-long
windows, then outputs a stream containing the twenty lowest Fourier
coefficients for each window as metrics and properties of the window
time (hour of day, month) as attributes. Placing this feature
transformation functionality at the start of the pipeline allows users
to encode domain-specific analyses without modifying later stages. The
base type of the stream is unchanged ($Point \rightarrow Point$),
allowing transforms to be chained. For specialized
data types like video frames, operators can subclass \textit{Point} to
further increase the specificity of types (e.g.,
\textit{VideoFramePoint}).

\minihead{3.) Classification} Following ingestion, \macrobase performs
classification, labeling each \textit{Point} according to its input
metrics. Both training and evaluating
classifiers on the metrics in the incoming data stream occur in this stage. \macrobase
supports a range of models, which we describe in
Section~\ref{sec:evaluation}. The simplest include rule-based models,
which check specific metrics for particular values (e.g.,
if the \texttt{Point} metric's L2-norm is greater than a fixed
constant). In Section~\ref{sec:detection}, we describe \macrobase's
default unsupervised models, which perform density-based
classification into ``outlier'' and ``inlier'' classes. Users can also
use operators that make use of supervised and pre-trained
models. Independent of model type, each classifier returns a stream of
labeled \textit{Point} outputs ($Point \rightarrow \texttt{(label,
  $Point$)}$).

\minihead{4.) Explanation} Rather than returning all labeled data
points, \macrobase aggregates the stream of labeled data points by
generating \textit{explanations}. As we describe in detail in
Section~\ref{sec:summarization}, \macrobase's default pipeline returns
explanations in the form of attribute-value combinations (e.g., device
ID $5052$) that are common among outlier points but uncommon among
inlier points. For example, at \cmt, \macrobase could highlight
devices that were found in at least $0.1\%$ of outlier trips and were
at least $3$ times more common among outliers than inliers. Each
explanation operator returns a stream of these aggregates
(\texttt{(label, $Point$)}$\rightarrow$\texttt{$Explanation$}), and
explanation operators can subclass $Explanation$ to provide additional
information, such as statistics about the explanation or
representative sequences of points to contextualize time-series
outliers.

Because \macrobase processes streaming data, explanation operators
continuously summarize the stream. However, continuously emitting
explanations may be wasteful if users only need
explanations at the granularity of seconds, minutes, or longer. As a
result, \macrobase's explanation operators are designed to emit explanations on
demand, either in response to a user request, or in response to a
periodic timer. In this way, explanation operators act as streaming view
maintainers.

\minihead{5.) Presentation} The number of output explanations may still be
large. As a result, most pipelines rank explanations by statistics specific to the explanations before
presentation. For example, by default,
\macrobase delivers a ranked list of explanations---sorted by their degree of outlier---occurrence to downstream consumers. 
\macrobase's default presentation mode is a static report
rendered via a REST API or GUI. In the former, programmatic
consumers (e.g., reporting tools such as PagerDuty) can automatically
forward explanations to downstream reporting or operational
systems. In the GUI, users can interactively inspect
explanations and iteratively define their \macrobase queries. In
practice, we have found that GUI-based exploration is an important
first step in formulating standing \macrobase queries that can later
be used in production.

\minihead{Extensibility} As we discussed in Section~\ref{sec:intro}
and demonstrate in Section~\ref{sec:casestudy}, \macrobase's pipeline
architecture lends itself to three major means of
extensibility. First, users can add new domain-specific feature
transformations to the start of a pipeline without modifying the rest of
the pipeline. Second, users can input rules and/or labels to \macrobase
to perform supervised classification. Third, users can write their own
feature transformation, classification, and explanation operators,
as well as new pipelines. This third option is the most labor-intensive, but is
also the interface with which \macrobase's maintainers author new
pipelines. These interfaces have proven useful to non-experts: a
master's student at Stanford and a master's student at MIT each
implemented and tested a new outlier detector operator in less than a
week of part-time work, and \macrobase's core maintainers currently
require less than an afternoon of work to author and test a new
pipeline.

By providing a set of interfaces with which to extend pipelines (with
varying expertise required), \macrobase places emphasis on ``pay as
you go'' deployment~\cite{mb-cidr}. \macrobase's Default Pipeline (\mdp, which we
illustrate in Figure~\ref{fig:arch-detail} and describe in the
following two sections) is optimized for efficient, accurate execution
over a variety of data types without relying on labeled data or
rules. It foregoes domain-specific feature extraction and instead
operates directly on raw input metrics. However, as we illustrate in
Section~\ref{sec:casestudy}, this interface design enables users to
incorporate more sophisticated features such as domain-specific
feature transformation, time-series analysis, and supervised models.

In this paper, we present \macrobase's interfaces using
an object-oriented interface, reflecting their current
implementation. However, each of \macrobase's operator types is
compatible with existing stream-to-relation semantics~\cite{cql},
theoretically allowing additional relational and stream-based
processing between stages. Realizing this mapping and the potential
for higher-level declarative interfaces above \macrobase's
 pipelines are promising areas for future work.

 \begin{figure} \includegraphics[width=\columnwidth]{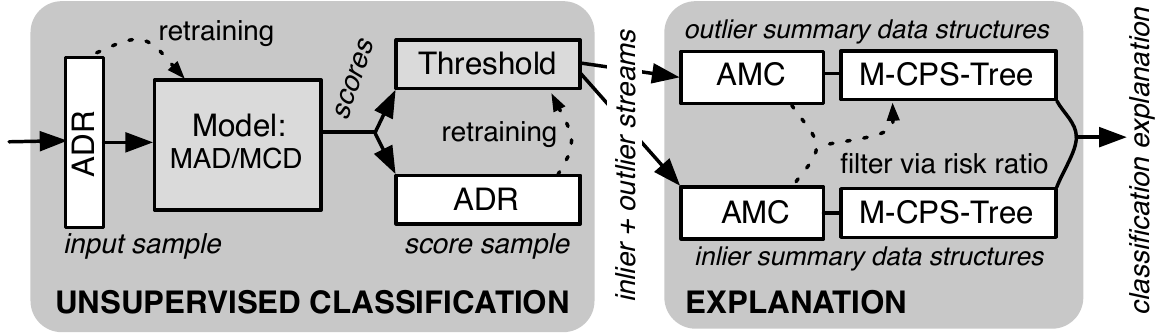} \caption{\mdp:
     \macrobase's default streaming classification
     (Section~\ref{sec:detection}) and explanation
     (Section~\ref{sec:summarization}) operators.}
\label{fig:arch-detail}
\end{figure}

\minihead{Operating modes} \macrobase supports three operating
modes. First, \macrobase's graphical front-end allows users to
interactively explore their data by configuring different inputs and
selecting different combinations of metrics and attributes. This is
typically the first step in interacting with the engine. Second,
\macrobase can execute {one-shot} queries that can be run
programmatically in a single pass over the data. Third, \macrobase can
execute {streaming} queries that can be run programmatically over
a potentially infinite stream of data. In streaming mode, \macrobase
continuously ingests data points and supports
exponentially decaying averages that give precedence to
more recent points (e.g., decreasing the importance of points at a
rate of $50\%$ every hour). \macrobase continuously re-renders query
results, and if desired, triggers automated alerting for downstream
consumers.

\vspace{1em}

\section{\mdp Classification}
\label{sec:detection}

\macrobase's classification operators label input data points, and, by default, identify data points that exhibit deviant behavior. While \macrobase allows users to configure their own operators, in this section, we focus on the design of \macrobase's default classification operators in \mdp, which use robust estimation procedures to fit a distribution to data streams and identify the least likely points with the distribution using quantile estimation. To enable streaming execution, we introduce the Adaptable Damped Reservoir, which \macrobase uses for model retraining and quantile estimation.

\subsection{Robust Distribution Estimation}

\mdp relies on unsupervised density-based classification to identify points that are abnormal relative to a population. However, a small number of anomalous points can have a large impact on density estimation. As an example, consider the \textit{Z-Score} of a point drawn from a univariate sample, which measures the number of standard deviations that the point lies away from the sample mean. This provides a normalized way to measure the ``outlying''-ness of a point (e.g., a Z-Score of three indicates the point lies three standard deviations from the mean). However, the Z-Score is not robust to outliers: a single outlying value can skew the mean and standard deviation by an unbounded amount, limiting its utility.

To address this challenge, \macrobase's \mdp pipeline leverages robust statistical estimation~\cite{robust-huber}, a branch of statistics that pertains to finding statistical distributions for data that is mostly well-behaved but may contain a number of ill-behaved data points. Given a distribution that reliably fits most of the data, we can measure each point's distance from this distribution in order to find outliers~\cite{robust-maronna}.

For univariate data, a robust variant of the Z-Score is to use the median and the Median Absolute Deviation (MAD), in place of mean and standard deviation, as measures of the location and scatter of the distribution. The MAD measures the median of the absolute distance from each point in the sample to the sample median. Since the median itself is resistant to outliers, each outlying data point has limited impact on the MAD score of all other points in the sample.

\begin{figure}
\includegraphics[width=\columnwidth]{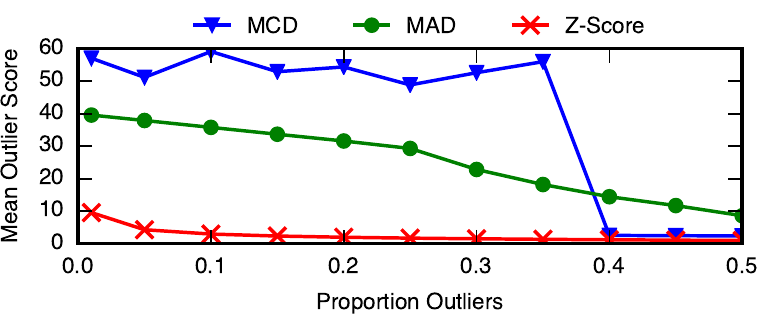}\vspace{-1em}
\caption{Discriminative power of estimators under 
  contamination by outliers (high scores better). Robust methods (MCD, MAD) outperform the Z-score-based approach.}
\label{fig:discriminatory}
\end{figure}

For multivariate data, the Minimum Covariance Determinant (MCD) provides similar robust estimates for location and spread~\cite{mcd-survey}. The MCD estimator finds the tightest group of points that best represents a sample, and summarizes the set of points according to its location $\mu$ and scatter $C$ (i.e., covariance) in metric space. Given these estimates, we can compute the distance between a point $x$  and the distribution via the \textit{Mahalanobis distance} $\sqrt{(x-\mu)^T C^{-1} (x-\mu)}$; intuitively, the Mahalanobis distance normalizes (or warps) the metric space via the scatter and then measures the distance to the center of the transformed space using the mean (see also Appendix~\ref{appendix:detect}).

As Figure~\ref{fig:discriminatory} empirically demonstrates, MAD and MCD  reliably identify points in outlier clusters despite increasing outlier {contamination} (experimental setup in Appendix~\ref{apdx:contamination}). Whereas MAD and MCD are resilient to contamination up to $50\%$, the Z-Score is unable to distinguish inliers and outliers under even modest contamination.

\minihead{Classifying outliers} Given a query with a single, univariate metric, \mdp uses a MAD-based detector, and, given a query with multiple metrics, \macrobase computes the MCD via an iterative approximation called FastMCD~\cite{fastmcd}.  These unsupervised models allow \mdp to score points without requiring labels or rules from users. Subsequently, \mdp uses a percentile-based cutoff over scores to identify the most extreme points in the sample. Points with scores above the percentile-based cutoff are classified as outliers, reflecting their distance from the body of the distribution.

As a notable caveat, MAD and MCD are parametric estimators, assigning scores based on an assumption that data is normally distributed. While extending these estimators to multi-modal behavior is straightforward~\cite{mcd-multicluster} and \macrobase allows substitution of more sophisticated detectors (e.g., Appendix~\ref{apdx:nonfree}), we do not consider them here. Instead, we have found that looking for far away points using these parametric estimators yields useful results: as we empirically demonstrate, many interesting behaviors manifest as extreme deviations from the overall population. Robustly locating the center of a population---while ignoring local, small-scale deviations in the body of the distribution---suffices to identify many important classes of outliers in the applications we study (cf.~\cite{mcd-distribution}).

\subsection{\mdp Streaming Execution}

Despite their utility, we are not aware of an existing algorithm for training MAD or MCD in a streaming context.\footnote{Specifically, MAD requires computing the median of median distances, meaning streaming quantile estimation alone is insufficient. FastMCD is an inherently iterative algorithm that iteratively re-sorts data.} This is especially problematic because, as the distributions within data streams change over time, \mdp's estimators should be updated to reflect the change.

\minihead{ADR: Adaptable Damped Reservoir} \mdp's solution to the retraining problem is a novel adaptation of reservoir sampling over streams, which we call the Adaptable Damped Reservoir (ADR). The ADR maintains a sample of input data that is exponentially weighted towards more recent points; the key difference from traditional reservoir sampling is that the ADR operates over \textit{arbitrary} window sizes, allowing greater flexibility than existing damped samplers. As Figure~\ref{fig:arch-detail} illustrates, \mdp maintains an ADR sample of the input to periodically recompute its robust estimator and a second ADR sample of the outlier scores to periodically recompute its quantile threshold.

The classic reservoir sampling technique can be used to select a uniform sample over a set of data using finite space and one pass~\cite{reservoirsample}. The probability of insertion into the sample, or ``reservoir,'' is inversely proportional to the number of points observed thus far. In the context of stream sampling, we can treat the stream as an infinitely long set of points and the reservoir as a uniform sample over the data observed so far.

In \macrobase, we wish to promptly reflect changes in the underlying data stream, and therefore we adapt a \textit{weighted} sampling approach, in which the probability of data retention decays over time. The literature contains several existing algorithms for weighted reservoir sampling~\cite{efraimidis-sampling,achao,aggarwal-sample}. Most recently, Aggarwal described how to perform exponentially weighted sampling on a per-record basis: that is, the probability of insertion is an exponentially weighted function of the number of points observed so far~\cite{aggarwal-sample}. While this is useful, as we demonstrate in Section~\ref{sec:evaluation}, under workloads with variable arrival rates, we may wish to employ a decay policy that decays in \textit{time}, not in number of tuples; specifically, tuple-at-a-time decay may skew the reservoir towards periods of high stream volume. 

To support more flexible reservoir behavior, \macrobase adapts an earlier variant of weighted reservoir sampling due to Chao~\cite{efraimidis-sampling, achao} to provide the first exponentially decayed reservoir sampler that decays over {arbitrary} decay intervals. We call this variant the \textit{Adaptable Damped Reservoir}, or ADR (Algorithm~\ref{alg:fdr}). In contrast with existing approaches that decay on a per-tuple basis, the ADR separates the insertion process from the decay decision, allowing both time-based and tuple-based decay policies. Specifically, the ADR maintains a running count $c_w$ of items inserted into the reservoir (of size $k$) so far. When an item is inserted, $c_w$ is incremented by one (or an arbitrary weight, if desired). With probability $\frac{k}{c_w}$, the item is placed into the reservoir and a random item is evicted from the reservoir. When the ADR is decayed (e.g., via a periodic timer or tuple count), its running count is multiplied by a decay factor (i.e., $c_w \coloneqq (1-\alpha)c_w$).

\begin{algorithm}[t]
\begin{algorithmic}
\State \textbf{given:} $k$: reservoir size $\in \mathbb{N}$; $r$: decay rate $\in (0,1)$
\State \textbf{initialization:} reservoir $R \gets \{\}$; current weight $ c_w \gets 0$

\Function{observe}{$x$: point, $w$: weight}
\State $c_w \gets c_w + w$
\If{$|R| < k$}
\State $R \gets R \cup \{x\}$
\Else{~with probability $\frac{k}{c_w}$}
\State remove random element from $R$ and add $x$ to $R$
\EndIf 
\EndFunction

\Function{decay}{ }
\State $c_w \gets r \cdot c_w$
\EndFunction
\end{algorithmic}
\caption{ADR: Adaptable Damped Reservoir}
\label{alg:fdr}
\end{algorithm}

\macrobase currently supports two decay policies: time-based decay, which decays the reservoir at a pre-specified rate measured according to real time, and batch-based decay, which decays the reservoir at a pre-specified rate measured by arbitrarily-sized batches of data points (Appendix~\ref{appendix:variable-fdr}). The validity of this procedure follows from Chao's sampler, which otherwise requires the user to manually manage weights and decay. As in Chao's sampler, in the event of extreme decay, ``overweight'' items with relative insertion probability $\frac{k}{c_w} > 1$ are always retained in the reservoir until their insertion probability falls below $1$, at which point they are inserted normally.

\macrobase's \mdp uses the ADR to solve the model retraining and quantile estimations problems:

\minihead{Maintaining training inputs} Either on a tuple-based or time-based interval, \mdp retrains models using the contents of an ADR that samples the input data stream. This streaming robust estimator maintenance and evaluation strategy is the first of which we are aware. We discuss this procedure's statistical impact in Appendix~\ref{apdx:samples}.

\minihead{Maintaining percentile thresholds} While streaming quantile estimation is well studied, we were not able to find many computationally inexpensive options for an exponentially damped model with arbitrary window sizes. Thus, instead, \macrobase uses an ADR to sample the outlier scores produced by MAD and MCD. The ADR maintains an exponentially damped sample of the scores, which it uses to periodically compute the appropriate score quantile value (e.g., 99th percentile of scores).\footnote{This enables a simple mechanism for detecting quantile drift: if the proportion of outlier points significantly deviates from the target percentile (i.e., via application of a binomial proportion confidence interval), \mdp should recompute the quantile.}   A sample of size $O(\frac{1}{\epsilon^2}\log(\frac{1}{\delta}))$ yields an $\epsilon$-approximation of an arbitrary quantile with probability $1-\delta$~\cite{quantiles}, so a ADR of size $20$K provides an $\epsilon=1\%$ approximation with $99\%$ probability ($\delta=1\%$).

\section{\mdp Explanation}
\label{sec:summarization}

\mdp's explanation operators produce explanations to contextualize and differentiate inliers and outliers according to their attributes. In this section, we discuss how \macrobase performs this task by using a metric from epidemiology, the \textit{relative risk ratio} (risk ratio), using a range of data structures. We again begin with a discussion of \mdp's batch-oriented operation and introduce a cardinality-based optimization, then discuss how \macrobase executes streaming explanation via the Amortized Maintenance Counter sketch.

\subsection{Semantics: Support and Risk Ratio}

\macrobase produces explanations that describe attributes common to outliers but relatively uncommon to inliers. To identify combinations of attribute values that are relatively common in outliers, \mdp finds combinations with high \textit{risk ratio (or relative risk ratio)}. This ratio is a standard diagnostic measure used in epidemiology, and is used to determine potential causes for disease~\cite{risk-confidence}. Formally, given an attribute combination appearing $a_o$ times in the outliers and $a_i$ times in the inliers, where there are $b_o$ other outliers and $b_i$ other inliers, the risk ratio is defined as: 
$$ \text{risk ratio} = \frac{a_o/(a_o + a_i)}{b_o/(b_o + b_i)}$$
Intuitively, the risk ratio quantifies how much more likely a data point is to be an outlier if it is of a specific attribute combination, as opposed to the general population.  To eliminate explanations corresponding to rare but non-systemic combinations, \mdp finds combinations with high \textit{support}, or occurrence (by relative count) in outliers. To facilitate these two tests, \mdp accepts a minimum risk ratio and level of outlier support as input parameters. As an example, $\mdp$ may find that $500$ of $890$ records flagged as outliers correspond to iPhone 6 devices (outlier support of $56.2\%$), but, if $80191$ of $90922$ records flagged as inliers also correspond to iPhone 6 devices (inlier support of $88.2\%$), we are likely uninterested in iPhone 6 as it has a low risk ratio of $0.1767$. \mdp reports explanations in the form of combinations of attributes, each subset of which has risk ratio and support above threshold.

\subsection{Basic Explanation Strategy}

A na\"{i}ve solution to computing the risk ratio for various attribute sets is to search twice, once over all inlier points and once over all outlier points, and then look for differences between the inlier and outlier sets. As we experimentally demonstrate in Section~\ref{sec:evaluation}, this is inefficient as it wastes times searching over attributes in inliers that are eventually filtered due to insufficient outlier support. Moreover, the number of outliers is much smaller than the inliers, so processing the two sets independently ignores the possibility of additional pruning. To reduce this wasted effort, \macrobase takes advantage of both the cardinality imbalance between inliers and outliers as well as the joint explanation of each set.

\begin{algorithm}[t]
\begin{algorithmic}[1]
\Statex \textbf{given:} minimum risk ratio $r$, minimum support $s$,\par
\hspace{0.8em} set of outliers $O$, set of inliers $I$
\State find attributes w/ support $\geq s$ in $O$ and risk ratio $\geq r$ in $O, I$ \label{l:fi}
\State mine FP-tree over $O$ using only attributes from (\ref{l:fi}) \label{l:fp}
\State filter (\ref{l:fp}) by removing patterns w/ risk ratio $< r$ in $I$; return
\end{algorithmic}
\caption{\mdp's Outlier-Aware Explanation Strategy}
\label{alg:summarize}
\end{algorithm}

\minihead{Optimization: Exploit cardinality imbalance} The cardinality of the outlier set is by definition much smaller than that of the inlier set. Therefore, instead of searching the outlier supports and the inlier supports separately, \mdp first finds outlier attribute sets with minimum support and subsequently searches the inlier attributes, while only searching for attributes that were supported in the outliers. This reduces the space of inlier attributes to explore.

\minihead{Optimization: Individual item ratios are cheap} We have found that many important attribute combinations (i.e., with high risk ratio) can be explained by a small number of attributes (typically, one or two, which can be tested inexpensively). Moreover, while computing risk ratios for all attribute combinations is expensive (combinatorial), computing risk ratios for single attributes is inexpensive: we can compute support counts over both inliers and outliers via a single pass over the attributes. Accordingly, \mdp first computes risk ratios for single attribute values, then computes support of combinations whose members have sufficient risk ratios.

In contrast with~\cite{ratio-itemset}, this optimization for risk ratio computation is enabled by the fact that we wish to find \textit{combinations} of attributes whose subsets are each supported and have minimum risk ratio. If a set of attributes is correlated, reporting them as a group helps avoid overwhelming the user with explanations. 

\minihead{Algorithms and Data Structures} In the one-pass batch setting, single attribute value counting is straightforward, requiring a single pass over the data; the streaming setting below is more interesting. We experimented with several itemset mining techniques that use dynamic programming to prune the search over attribute combinations with sufficient support and ultimately decided on prefix-tree-based approaches inspired by FPGrowth~\cite{fpgrowth}. In brief, the FPGrowth algorithm maintains a frequency-descending prefix tree of attributes that can subsequently be mined by recursively generating a set of ``conditional'' trees. Corroborating recent benchmarks~\cite{spmf-lib}, the FPGrowth algorithm was fast and proved extensible in our streaming implementation below.

\minihead{End result} The result is a three-stage process (Algorithm~\ref{alg:summarize}). \mdp first calculates the attribute values with minimum risk ratio (support counting, followed by a filtering pass based on risk ratio). From the first stage's outlier attribute values, \mdp then computes supported outlier attribute combinations. Finally, \mdp computes the risk ratio for each attribute combination based on their support in the inliers (support counting, followed by a filtering pass to exclude any attribute combinations with insufficient risk ratio).

\minihead{Significance} We discuss confidence intervals on \mdp explanations as well as quality improvements achievable by processing large data volumes in Appendix~\ref{appendix:confidence}.

\subsection{Streaming Explanation}

As in \mdp detection, streaming explanation generation is more challenging. We present the \mdp implementation of single-attribute streaming explanation then extend the approach to multi-attribute streaming explanation.

\minihead{Implementation: Single Attribute Summarization} To begin, we find individual attributes with sufficient support and risk ratio while respecting both changes in the stream and limiting the overall amount of memory required to store support counts. The problem of maintaining a count of frequent items (i.e., \textit{heavy hitters}, or attributes with top $k$ occurrence) in data streams is well studied~\cite{cormode-survey}. Given a heavy-hitters sketch over the inlier and outlier stream, we can compute an approximate support and risk ratio for each attribute by comparing the contents of the sketches at any time.

Initially, we implemented the \mdp item counter using the SpaceSaving algorithm~\cite{spacesaving}, which provides empirically good performance~\cite{cormode-item-survey} and has extensions in the exponentially decayed setting~\cite{cormode-decay}. However, like many of the sketches in the literature, SpaceSaving was designed to strike a balance between sketch size and performance, with a strong emphasis on limited size. For example, in its heap-based variant, SpaceSaving maintains $\frac{1}{k}$-approximate counts for the top $k$ item counts by maintaining a heap of the items. For a stream of size $n$, this requires $O(n\log(k))$ update time. (In the case of exponential decay, the linked-list variant can require $O(n^2)$ processing time.)

While logarithmic update time is modest for small sketches, given only two heavy-hitters sketches per \macrobase query, \mdp can expend more memory on its sketches to improve accuracy; for example, 1M items require four megabytes of memory for float-encoded counts, which is small relative to modern server memory sizes. As a result, we developed a heavy-hitters sketch, called the \textit{Amortized Maintenance Counter} (AMC, Algorithm~\ref{alg:amc}), that occupies the opposite end of the design spectrum: the AMC uses a much greater amount of memory for a given accuracy level, but is faster to update and still limits total space utilization. The key insight behind the AMC is that if we observe even a single item in the stream more than once, we can amortize the overhead of maintaining the sketch across multiple observations of the same item. In contrast, SpaceSaving maintains the sketch for every observation but in turn ensures a smaller sketch size.

\begin{algorithm}[t]
\begin{algorithmic}
\State \textbf{given:} $\epsilon \in (0, 1)$; $r$: decay rate $\in (0,1)$
\State \textbf{initialization:} $C$ (item $\rightarrow$ count) $\gets \{\}$; weight $ w_i \gets 0$

\Function{observe}{$i$: item, $c$: count}
\State $C[i] \gets w_i+c$ if $i \notin C$ else $C[i]+c$
\EndFunction

\Function{maintain}{ }
\State remove all but the $\frac{1}{\epsilon}$ largest entries from $C$
\State $w_i \gets $ the largest value just removed, or, if none removed, $0$
\EndFunction

\Function{decay}{ }
\State decay the value of all entries of $C$ by $r$
\State call \textsc{maintain}(~)
\EndFunction
\end{algorithmic}
\caption{AMC: Amortized Maintenance Counter}
\label{alg:edafic}
\label{alg:amc}
\end{algorithm}

AMC provides the same counting functionality as a traditional heavy-hitters sketch but exposes a second method, \textit{maintain}, that is called to periodically prune the sketch size. AMC allows the sketch size to increase between calls to \textit{maintain}, and, during maintenance, the sketch size is reduced to a desired stable size, which is specified as an input parameter. Therefore, the maximum size of the sketch is controlled by the period between calls to \textit{maintain}: as in SpaceSaving, a stable size of $\frac{1}{\epsilon}$ yields an $n\epsilon$ approximation of the count of $n$ points, but the size of the sketch may grow within a period. This separation of insertion and maintenance has two implications. First, it allows constant-time insertion, which we describe below. Second, it allows a range of maintenance policies, including a sized-based policy, which performs maintenance once the sketch reaches a pre-specified upper bound, as well as a variable period policy, which operates over real-time and tuple-based windows (similar to ADR).

To implement this functionality, AMC maintains a set of approximate counts for all items that were among the most common in the previous period along with approximate counts for all other items that observed in the current period. During maintenance, AMC prunes all but the $\frac{1}{\epsilon}$ items with highest counts and records the maximum count that is discarded ($w_i$). Upon insertion, AMC checks to see if the item is already stored. If so, the item's count is incremented. If not, AMC stores the item count plus $w_i$. If an item is not stored in the current window, the item must have had count less than or equal to $w_i$ at the end of the previous period.

AMC has three major differences compared to SpaceSaving. First, AMC updates are constant time (hash table insertion) compared to $O(\log(\frac{1}{\epsilon}))$ for SpaceSaving. Second, AMC has an additional maintenance step, which is amortized across all items seen in a window. Using a min-heap, with $I$ items in the sketch, maintenance requires $O(I\cdot\log(\frac{1}{\epsilon}))$ time. If we observe even one item more than once, this is faster than performing maintenance on every observation.  Third, AMC has higher space overhead; in the limit, it must maintain all items it has seen between maintenance intervals.

\minihead{Implementation: Streaming Combinations} While AMC tracks single items, \mdp also needs to track combinations of attributes. As such, we sought a tree-based technique that would admit exponentially damped arbitrary windows but eliminate the requirement that each attribute be stored in the tree, as in recent proposals such as the CPS-tree~\cite{cps-tree}. As a result, \mdp adapts a combination of two data structures: AMC for the frequent attributes, and an adaptation of the CPS-Tree data structure to store frequent attributes. We present algorithms for maintaining the adapted CPS-tree in Appendix~\ref{appendix:confidence}.



\minihead{Summary} \mdp's streaming explanation operator consists of two primary parts: maintenance and querying.  When a new data point arrives at the  summarization operator, \macrobase inserts each of the point's attributes into an AMC sketch.  \macrobase then inserts a subset of the point's attributes into a prefix tree that maintains an approximate, frequency descending order. When a window has elapsed, \macrobase decays the counts of the items and the counts in each node of the prefix tree. \macrobase removes any attributes that are no longer above the support threshold and rearranges the prefix tree in frequency-descending order. To produce explanations, \macrobase runs FPGrowth on the prefix tree.

\section{Evaluation}
\label{sec:evaluation}

In this section, we evaluate the accuracy, efficiency, and flexibility
of \macrobase and the \mdp operators. We wish to
demonstrate that:

\begin{itemize}[itemsep=.1em,parsep=.25em,topsep=.25em]
\item \macrobase is accurate: on controlled, synthetic
  data, under changes in stream behavior and over real-world
  workloads from the literature and in production
  (Section~\ref{sec:accuracy}).

\item \macrobase can process up to $2$M points per second per query on
  a range of real-world datasets (Section~\ref{sec:speed}).

\item \macrobase's cardinality-aware explanation strategy produces
  meaningful speedups (average: 3.2$\times$ speedup;
  Section~\ref{sec:microbenchmarks}).

\item \macrobase's use of AMC is up to 500$\times$ faster than existing
  sketches on production data (Section~\ref{sec:comparison}).

\item \macrobase's architecture is extensible, which we illustrate via
  three case studies (Section~\ref{sec:casestudy}).
\end{itemize}

\minihead{Experimental environment} We report results from deploying
the \macrobase prototype on a server with four Intel Xeon E5-4657L
2.40GHz CPUs containing 12 cores per CPU and 1TB of RAM. To isolate
the effects of pipeline processing, we exclude loading time from our
results. By default, we issue \mdp queries with a minimum support of
$0.1\%$ and minimum risk ratio of $3$, a target outlier percentile of
$1\%$, ADR and AMC sizes of 10K, a decay rate of 0.01 every 100K
points, and report the average of at least three runs per
experiment. We vary these parameters in subsequent experiments in this
section and the Appendix.

\minihead{Implementation} We describe \macrobase's implementation,
dataflow runtime, and approach to parallelism in
Appendix~\ref{sec:implementation}.

\minihead{Large-scale datasets} To compare the efficiency of \macrobase and
related techniques, we compiled a set of large-scale real-world
datasets (Table~\ref{table:queries}) for evaluation (descriptions in
Appendix~\ref{appendix:datasets}).

\subsection{Result Quality}
\label{sec:accuracy}

In this section, we focus on \macrobase's statistical result
quality. We evaluate precision/recall on synthetic and real-world
data, demonstrate adaptivity to changes in data streams, and report on
experiences from production usage.

\minihead{Synthetic dataset accuracy} We ran \mdp over a synthetic
dataset generated in line with those used to evaluate recent anomaly
detection systems~\cite{scorpion, perfaugur}. The generated dataset
contains 1M data points from a number of synthetic devices. Each device
in the dataset has a unique device ID attribute and metrics which are drawn from
either an inlier distribution ($\mathcal{N}(10, 10)$) or outlier
distribution ($\mathcal{N}(70, 10)$). We subsequently evaluated
\macrobase's ability to automatically determine the device IDs
corresponding to the outlying distribution. We report the $F_1$-score
$\left(2 \cdot~\frac{precision~\cdot~recall}{precision +
    recall}\right)$ for the set of device IDs identified as outliers
metric for explanation quality.
  
Since \mdp's statistical techniques are a natural match for this
experimental setup, we also perturbed the base experiment to understand when 
\mdp might underperform. We introduced two types of noise into the
measurements to quantify their effects on \mdp's performance. 
First, we introduced \textit{label noise} by randomly assigning readings from the
outlier distribution to inlying devices and vice-versa. Second, we introduced 
\textit{measurement noise} by randomly assigning a proportion of both outlying 
and inlying points to a third, uniform distribution over the interval $[0,80]$.

Figure~\ref{fig:precision-recall} illustrates the results.  In the the
noiseless regions of Figure~\ref{fig:precision-recall}, \mdp correctly
identified 100\% of the outlying devices. As the outlying devices are
solely drawn from the outlier distribution, constructing outlier
explanations via the risk ratio enables \macrobase to perfectly
recover the outlying device IDs. In contrast, techniques that rely
solely on individual outlier classification deliver less accurate
results on this workload (cf.~\cite{scorpion,perfaugur}). Under label
noise, \macrobase robustly identified the outlying devices until
approximately 25\% noise, which corresponds a $3:1$ ratio of correct
to incorrect labels. As our risk ratio threshold is set to $3$,
exceeding this threshold causes rapid performance degradation. Under
measurement noise, accuracy degrades linearly with the amount of
noise. \mdp is more robust to this type of noise when fewer devices
are present; its accuracy suffers with a larger number of devices, as
each device type is subject to more noisy readings.

\begin{figure}
\centering
\includegraphics[width=.9\columnwidth]{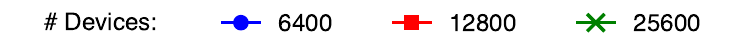}\\\vspace{-.5em}
\includegraphics[width=.48\linewidth]{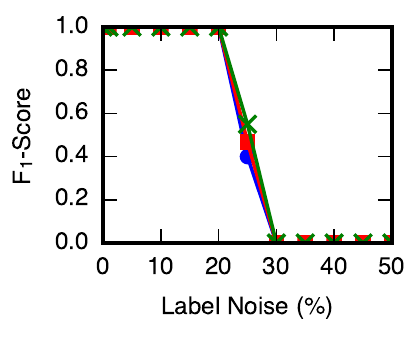} \hfill
\includegraphics[width=.48\linewidth]{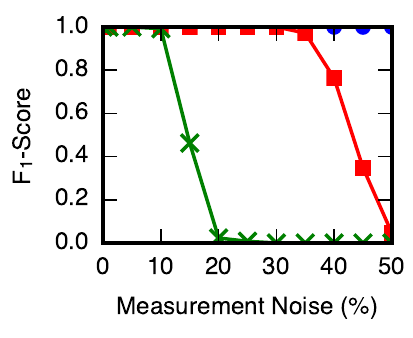} \vspace{-.5em}
\caption{Precision-recall of explanations. Without noise, \mdp
  exactly identifies misbehaving devices. \mdp's use of risk ratio
  improves resiliency to both label and measurement noise.}
\label{fig:precision-recall}
\end{figure}

In summary, \mdp is able to accurately identify correlated causes of
outlying data for noise of 20\% or more. The noise threshold is
improved by both \mdp's use of robust methods as well as the use of
risk ratio to prune irrelevant summaries. Noise of this magnitude is
likely rare in practice, and, if such noise exists, is possibly of another interesting behavior in the data.

\minihead{Real-world dataset accuracy} In addition to synthetic data,
we also performed experiments to determine \macrobase's ability to
accurately identify systemic abnormalities in real-world data. We
evaluated \macrobase's ability to distinguish abnormally-behaving OLTP
servers within a cluster, as defined according to data and manual
labels collected in a recent study~\cite{dbsherlock} to diagnose
performance issues within a {single} host. We performed a set of
experiments, each corresponding to a distinct type of performance
degradation within MySQL on a particular OLTP workload (TPC-C and
TPC-E). For each experiment, we consider a cluster of eleven servers,
where a single server exhibits the degradation. Using over 200
operating systems and database performance counters, we ran \mdp to
identify the anomalous server.

We ran \mdp with two sets of queries. In the former set, QS, \mdp
executed a query to find abnormal hosts (with hostname attributes)
using a {single} set of 15 metrics identified via feature selection
techniques on a holdout of 2 clusters per experiment (i.e., one query
for all anomalies). As Table~\ref{table:dbsherlock}
(Appendix~\ref{appendix:dbsherlock}) shows, under QS, \mdp achieves
top-1 accuracy of 86.1\% on the holdout set across all forms of
anomalies (top-3: 88.8\%). For eight of nine anomalies, \mdp's top-1
accuracy is higher: 93.8\%. However, for the ninth anomaly, which
corresponds to a poorly written query, the metrics correlated with the
anomalous behavior are substantially different.

In the second set of experiments, QE, \mdp executed a slow-hosts query
using a set of metrics for {each} distinct anomaly type (e.g., network
contention), again using a holdout of 2 clusters per experiment (i.e.,
one query per anomaly type). In contrast with QS, because QE targets
each type of performance degradation with a custom set of metrics, it
is able to identify behaviors more reliably, leading to perfect top-3
accuracy.

These results show that with proper feature selection,
\macrobase accurately recovers systemic causes even in
unsupervised settings.

\minihead{Adaptivity} While the previous set of experiments operated
over data with a static underlying distribution, we sought to understand the benefit 
of \mdp's ability to adapt to changes in the input distribution via the
exponential decay of ADR and AMC. We performed a controlled experiment over 
two types of time-varying behavior: changing underlying data distribution,
and variable data arrival rate. We then compared the accuracy of \mdp 
outlier detection across three sampling techniques: a uniform reservoir sample, 
a per-tuple exponentially decaying reservoir sample, and our proposed ADR.

\floatsetup[figure]{style=plain,subcapbesideposition=top}
\begin{figure}
\sidesubfloat[ ] {\label{fig:a}\includegraphics[width=0.94\columnwidth]{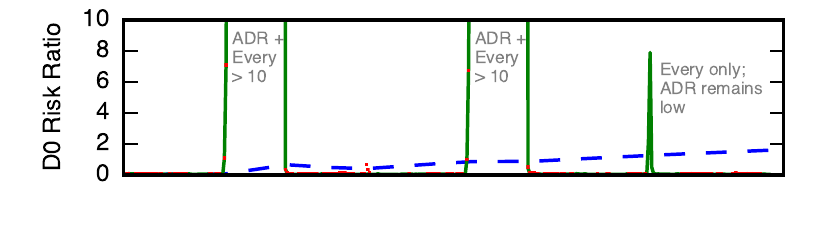}}\\[-3ex]
\sidesubfloat[ ] {\label{fig:b}\includegraphics[width=0.94\columnwidth]{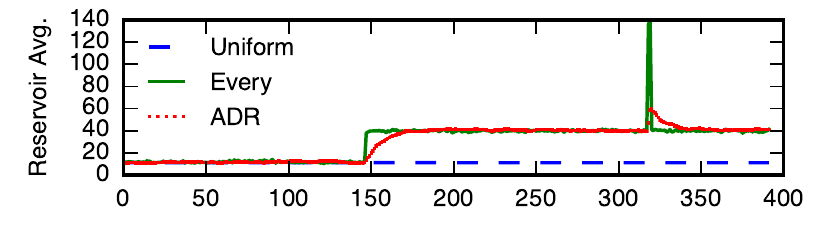}}\\[-3ex]
\sidesubfloat[ ] {\label{fig:c}\includegraphics[width=0.94\columnwidth]{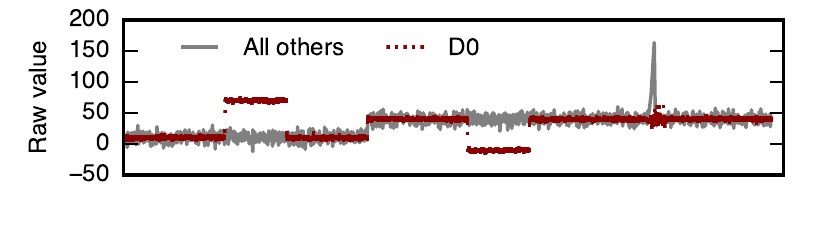}}\\[-3ex]
\sidesubfloat[ ]{\label{fig:d}\includegraphics[width=0.94\columnwidth]{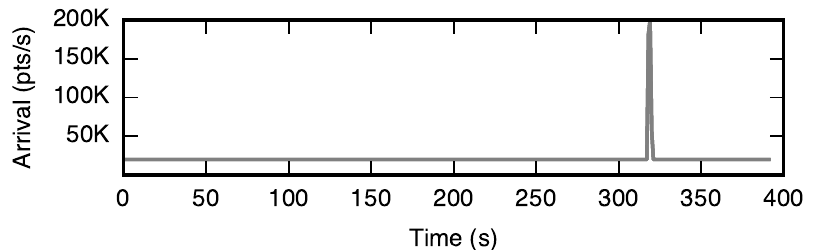}}
\caption{ ADR provides greater adaptivity compared to tuple-at-a-time
  reservoir sampling and is more resilient to spikes in data volume
  (see text for details).}
\label{fig:adaptivity}
\end{figure}
\clearfloatsetup{figure}

Figure~\ref{fig:c} displays the time-evolving stream representing 
100 devices over which \mdp operates. To begin, all devices
produce readings drawn from a Gaussian $\mathcal{N}(10, 10)$
distribution. After $50$ seconds, a single device, $D0$, produces
readings from $\mathcal{N}(70, 10)$ before returning to the original
distribution at $100$ seconds. The second period (150s to 300s) is
similar to the first, except we also introduce a shift in all devices'
metrics: after 150 seconds, all devices produce readings from
$\mathcal{N}(40, 10)$, and, after $225$ seconds, $D0$ produces
readings from $ \mathcal{N}(-10, 10)$, returning to $\mathcal{N}(40,
10)$ after $250$ seconds. Finally from 300s to 400s, all devices
experience a spike in data arrival rate. We introduce a four-second
noise spike in the sensor readings at 320 seconds: the arrival rate
rises by ten-fold, to over $200$k points per second, with
corresponding values drawn from a $\mathcal{N}(85, 15)$ distribution
(Figure~\ref{fig:d}).

In the first time period, all three strategies detect $D0$ as an
outlier, as reflected in the computed risk ratios in Figure~\ref{fig:a}. After
$100$ seconds, when $D0$ returns to the inlier distribution, its
risk ratio drops. The reservoir averages remain unchanged in all
strategies (Figure~\ref{fig:b}). In the second time period, both
adaptive reservoirs adjust to the new distribution by $170$ seconds,
while the uniform reservoir fails to adapt quickly
(Figure~\ref{fig:b}). As such, when $D0$ drops to $\mathcal{N}(-10,
10)$ from time $225$ through $250$, only the two adaptive strategies
track the change (Figure~\ref{fig:a}). At time $300$, the short noise
spike appears in the sensor readings. The per-tuple reservoir is
forced to absorb this noise, and the distribution in this reservoir
spikes precipitously. As a result, $D0$, which remains at
$\mathcal{N}(40, 10)$ is falsely suspected as outlying. In contrast,
the ADR average value rises slightly but never suspects $D0$ as an
outlier. This illustrates the value of \mdp's adaptivity to distribution
changes and resilience to variable arrival rates.

\minihead{Production results} \macrobase currently operates over a
range of production data and external users report the prototype has
discovered previously unknown and sometimes serious behaviors in
several domains. Here, we report on our experiences deploying
\macrobase at \cmt, where it identified several previously unknown
behaviors. In one case, \macrobase highlighted a small number of users
who experienced issues with their trip detection. In another case,
\macrobase discovered a rare issue with the \cmt application and a
device-specific battery problem. Consultation and investigation with
the \cmt team confirmed these issues as previously unknown, and have
since been addressed. These experiences and others~\cite{mb-cidr} have
proven a useful demonstration of \macrobase's ability to prioritize
attention in production environments and inspired several ongoing
extensions (Section~\ref{sec:conclusion}).

\subsection{End-to-End Performance}
\label{sec:speed}

In this section, we evaluate \macrobase's end-to-end performance on
real-world datasets. For each dataset $X$, we execute two \macrobase
queries: a simple query, with a single attribute and metric (denoted
$XS$), and a complex query, with a larger set of attributes and, when
available, multiple metrics (denoted $XC$).  We then report throughput
for two system configurations: one-shot batch execution that processes each
stage in sequence and exponentially-weighted streaming execution (EWS)
that processes points continuously. One-shot and EWS have different
semantics, as reflected in the explanations they produce. One-shot execution examines the entire dataset at once. Exponentially
weighted streaming prioritizes recent points. Therefore, for
datasets with few distinct attribute values (e.g., \accidents
contains only nine types of weather conditions), the explanations will
have high similarity. However, explanations differ in datasets with many distinct
attribute values (typically the complex queries with hundreds
of thousands of possible combinations---e.g., \disburse has 138,338
different disbursement recipients). For this reason, we provide throughput results both with and
without explanations, as well as the number of explanations generated
by the simple ($XS$) and complex ($XC$) queries and their Jaccard
similarity.

\begin{table*}
\center
\footnotesize
\setlength\tabcolsep{6 pt}
\footnotesize
\begin{tabular}{|c l c c l | r r |  r  r | r r | c | }
\hline
\multicolumn{5}{|c|}{\textbf{Queries}} & \multicolumn{2}{c|}{Thru w/o
  Explain (pts/s)} &
\multicolumn{2}{c|}{Thru w/ Explain (pts/s)} & \multicolumn{2}{c|}{\# Explanations}
& \multicolumn{1}{c|}{Jaccard} \\
Dataset & Name & {Metrics} & {Attrs} & { Points} & One-shot &
EWS & One-shot & EWS & One-shot & EWS & \multicolumn{1}{c|}{Similarity} \\ \hline


\multirow{2}{*}{Liquor} & LS &  1 & 1 & \multirow{2}{*}{3.05M} & 1549.7K & 967.6K & 1053.3K & 966.5K & 28 & 33 & 0.74 \\
& LC & 2 & 4 &  & 385.9K & 504.5K & 270.3K & 500.9K & 500 & 334 & 0.35 \\\hline
 \multirow{2}{*}{Telecom}  & TS & 1 & 1 & \multirow{2}{*}{10M}  & 2317.9K & 698.5K & 360.7K & 698.0K & 469 & 1 & 0.00 \\
& TC & 5 & 2 &  & 208.2K & 380.9K & 178.3K & 380.8K & 675 & 1 & 0.00 \\\hline
\multirow{2}{*}{Campaign} & ES &  1 & 1 & \multirow{2}{*}{10M} & 2579.0K & 778.8K & 1784.6K & 778.6K & 2 & 2 & 0.67 \\
& EC & 1 & 5 &  & 2426.9K & 252.5K & 618.5K & 252.1K & 22 & 19 & 0.17 \\\hline
\multirow{2}{*}{Accidents}  & AS &  1 & 1 & \multirow{2}{*}{430K} & 998.1K & 786.0K & 729.8K & 784.3K & 2 & 2 & 1.00 \\
& AC & 3 & 3 &  & 349.9K & 417.8K & 259.0K & 413.4K & 25 & 20 & 0.55 \\\hline
 \multirow{2}{*}{Disburse} & FS & 1 & 1 & \multirow{2}{*}{3.48M} & 1879.6K & 1209.9K & 1325.8K & 1207.8K & 41 & 38 & 0.84 \\
& FC  & 1 & 6 & & 1843.4K & 346.7K & 565.3K & 344.9K & 1710 & 153 & 0.05 \\\hline
\multirow{2}{*}{\cmtd} & MS & 1 & 1 & \multirow{2}{*}{10M} & 1958.6K & 564.7K & 354.7K & 562.6K & 46 & 53 & 0.63 \\
& MC & 7 & 6 &  & 182.6K & 278.3K & 147.9K & 278.1K & 255 & 98 & 0.29 \\\hline


\end{tabular}
\caption{Datasets and query names, throughput, and explanations produced under one-shot and exponentially weighted streaming (EWS)
  execution. \macrobase sustains throughput of several hundred thousand (and up to $2.5$M) points per second.}
\label{table:queries}
\end{table*}

Table~\ref{table:queries} displays results across all
queries. Throughput varied from 147K points per second (on $MC$ with
explanation) to over 2.5M points per second (on $TS$ without
explanation); the average throughput for one-shot execution was 1.39M
points per second, and the average throughput for EWS was 599K points
per second. The better-performing mode depended heavily on the
particular data set and characteristics. In general, queries with
multiple metrics were slower in one-shot than queries with single
metrics (due to increased training time, as streaming trains over
samples), and EWS typically returned fewer explanations due to its
temporal bias. Generating each explanation at the end of the query
incurred an approximately 22\% overhead. In all cases, these queries
far exceed the current arrival rate of data for each dataset. In
practice, users tune their decay on a per-application basis (e.g., at
\cmt, streaming queries may prioritize trips from the last hour to
catch errors arising from the most recent deployment). These
throughputs exceed those of related techniques we have encountered in
the literature (by up to three orders of magnitude); we examine
specific factors that contribute to this performance in the next
section.

\minihead{Runtime breakdown} To further understand how each pipeline
operator contributed to overall performance, we profiled \macrobase's
one-shot execution (EWS was challenging to instrument accurately
due to its streaming execution). On $MC$, \macrobase spent
approximately 52\% of its execution training MCD, 21\% scoring points,
and 26\% generating explanations. On $MS$, \macrobase spent approximately
54\% of its execution training MAD, 16\% scoring points, and 29\%
generating explanations. In contrast, on $FC$, which returned over 1000
explanations, \macrobase spent 31\% of its execution training MAD, 4\%
scoring points, and 65\% generating explanations. Thus, the overhead of each
component is data- and query-dependent.

\subsection{Microbenchmarks and Comparison}
\label{sec:microbenchmarks}
\label{sec:comparison}

In this section, we explore two key aspects of \macrobase's design:
cardinality-aware explanation and use of AMC sketches. 

\minihead{Cardinality-aware explanation} We evaluated the efficiency
of \macrobase's cardinality-aware pruning compared to traditional
FPGrowth. \macrobase leverages a unique pruning strategy that exploits
the low cardinality of outliers, which delivers large speedups---on
average, over 3$\times$ compared to unoptimized
FPGrowth. Specifically, \macrobase's produced a summary of each
dataset's inliers and outliers in 0.22--1.4 seconds. In contrast,
running FPGrowth separately on inliers and outliers was, on average,
3.2$\times$ slower; compared to \macrobase's joint explanation
according to support and risk ratio, much of the time spent mining
inliers (with insufficient risk ratio) in FPGrowth is wasted. However,
both \macrobase and FPGrowth must perform a linear pass over all of
the inliers, which places a lower bound on the running time. The 
benefit of this optimization depends on the risk ratio, which we vary in Appendix~\ref{apdx:support-oi}.

\minihead{AMC Comparison} We also compared the performance of AMC with
existing heavy-hitters sketches
(Figure~\ref{fig:itemcount-compare}). AMC outperformed both
implementations of SpaceSaving in all configurations by a margin of up
to 500$\times$ for sketch sizes exceeding 100 items. This is
because the SpaceSaving overhead (heap maintenance on every operation
is expensive with even modestly-sized sketches or list traversal is
costly for decayed, non-integer counts) is costly. In contrast, with
an update period of 10K points, AMC sustained over 10M updates
per second. The primary cost of these performance improvements is
additional space: for example, with a minimum sketch size of 10 items
and update period of 10K points, AMC retained up to 10,010 items while
each SpaceSaving sketch retained only 10. As a result, when memory
sizes are especially constrained, SpaceSaving may be preferable, at a
measurable cost to performance.

\begin{figure}
\centering
\includegraphics[width=.6\columnwidth]{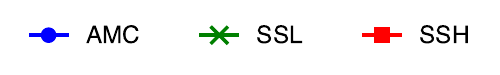}\\\vspace{-.5em}
\includegraphics[width=.48\linewidth]{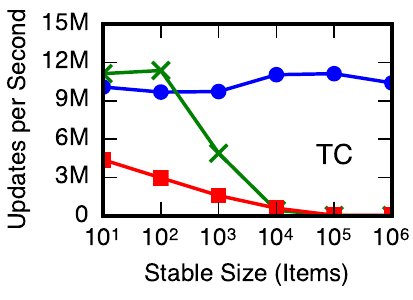} \hfill
\includegraphics[width=.48\linewidth]{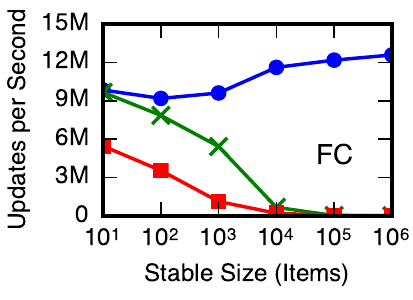}\vspace{-.5em}
\caption{Streaming heavy hitters sketch comparison. AMC: Amortized
  Maintenance Counter with maintenance every 10K items; SSL: Space
  Saving List; SSH: Space Saving Hash. All share the same accuracy
  bound. Varying the AMC maintenance period produced similar results.}
\label{fig:itemcount-compare}
\end{figure}

\minihead{Additional results} In Appendix~\ref{apdx:expts}, we provide
additional results examining the distribution of outlier scores, the
effect of varying support and risk ratio, the effect of training over
samples and operating over varying metric dimensions, the
behavior of the M-CPS tree, preliminary scale-out behavior, comparing
the runtime of \mdp explanation to both existing batch explanation
procedures, and \mdp detection and explanation to operators from
frameworks including Weka, Elki, and RapidMiner.

\subsection{Case Studies and Extensibility}
\label{sec:casestudy}
\label{sec:extensibility}
\let\oldcenter\center
\let\oldendcenter\endcenter
\renewenvironment{center}{\setlength\topsep{2pt}\oldcenter}{\oldendcenter}

\macrobase is designed for extensibility, as we highlight via
case studies in three separate domains. We describe the pipeline
structures, performance, and interesting explanations from applying
\macrobase over supervised, time-series, and video
surveillance data.

\minihead{Hybrid Supervision} We demonstrate \macrobase's ability to combine supervised and unsupervised classification models via a use case from \cmt. 
Each trip in the \cmtd dataset is accompanied by a supervised diagnostic score representing
the trip quality. While \mdp's unsupervised operators can use this score as
an input, \cmt also wishes to capture low-quality scores
independent of their distribution in the population. Accordingly, we
authored a new \macrobase pipeline that feeds some metrics (e.g. trip length, battery drain) to the \mdp
MCD operator and {also} feeds the diagnostic metric (trip quality score) to a
special rule-based operator that flags low quality scores as
anomalies. The pipeline, which we depict below, performs a logical \textit{or} over the two
classification results:
\begin{center}
\includegraphics[width=.8\columnwidth]{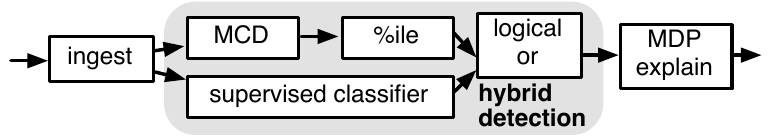}
\end{center}\vspace{-.25em}
With this hybrid supervision strategy, \macrobase identified additional
behaviors within the \cmtd dataset. Since the quality scores were
generated external to \macrobase and the supervision rule in
\macrobase was lightweight, runtime was unaffected. This kind of
pipeline can easily be extended to more complex supervised models.

\minihead{Time-series} \macrobase can also detect temporal behaviors via feature transformation, which we demonstrate using a dataset of 16M points
capturing a month of electricity usage from devices within a
household~\cite{power-dataset}. We augment \mdp by adding a sequence
of feature transforms that $i.)$ partition the stream by device ID,
$ii.)$ window the stream into hourly intervals, with attributes
according to hour of day, day of week, and date, then $iii.)$ apply a
Discrete-Time Short-Term Fourier Transform (STFT) to each window, and
truncate the transformed data to a fixed number of dimensions. As the
diagram below shows, we feed the transformed stream into an unmodified
\mdp and search for outlying time periods and devices:
\begin{center}
\includegraphics[width=.91\columnwidth]{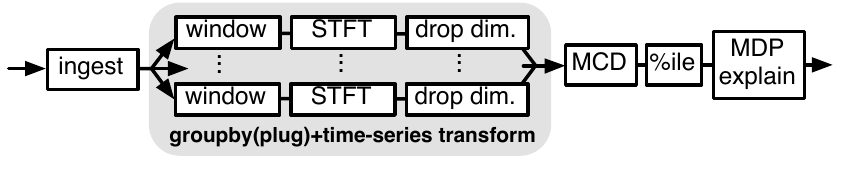}
\end{center}\vspace{-.25em}
With this custom time-series pipeline, \macrobase detected several systemic periods
of abnormal device behavior. For example, the following dataset of power usage by a household refrigerator
 spiked on an hourly basis (possibly corresponding to
compressor activity); instead of highlighting the hourly power spikes, \macrobase was able to detect that the refrigerator
consistently behaved abnormally compared to other devices in
the household and to other time periods between the hours of
12PM and 1PM---presumably, lunchtime---as highlighted in the excerpt
below:
\begin{center}
\includegraphics[width=\columnwidth]{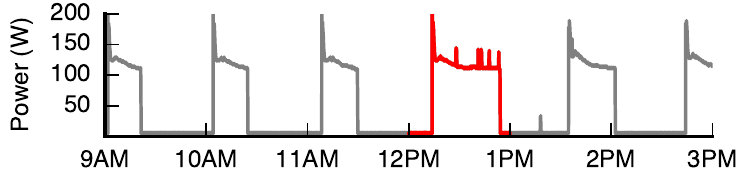}
\end{center}
Without feature transformation, the entire \mdp pipeline completed in
158ms.  Feature transformation dominated the runtime, utilizing 516
seconds to transform the 16M points via unoptimized STFT.

\minihead{Video Surveillance} We further highlight \macrobase's ability to easily
operate over a wide array of data sources and domains by searching for interesting patterns in the CAVIAR video surveillance dataset~\cite{caviar}. Using OpenCV 3.1.0, we add a custom feature transform that computes the average optical flow velocity between video frames, a technique that has been successfully applied in human action detection~\cite{fightdetection}. Each transformed frame is tagged with a
time interval attribute, which we use to identify interesting video
segments and, as depicted below, the remainder of the pipeline
executes the standard \mdp operators:
\begin{center}
\includegraphics[width=.8\columnwidth]{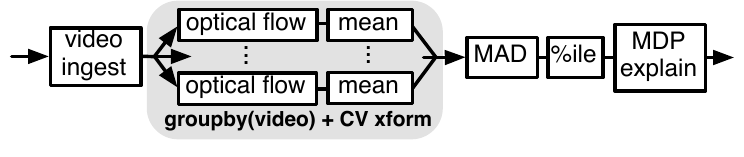}
\end{center}\vspace{-.25em}
Using this pipeline, \macrobase detected periods of abnormal motion in
the video dataset. For example, the \macrobase pipeline highlighted
a three-second period in which two people fought:
\begin{center}
\includegraphics[width=.18\columnwidth]{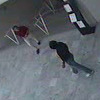}\hfill
\includegraphics[width=.18\columnwidth]{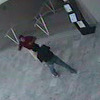}\hfill
\includegraphics[width=.18\columnwidth]{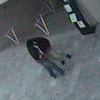}\hfill
\includegraphics[width=.18\columnwidth]{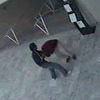}\hfill
\includegraphics[width=.18\columnwidth]{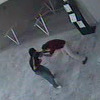}\hfill
\end{center}\vspace{-.5em}
Like our STFT pipeline, feature transformation via
optical flow dominated runtime ($22$s vs.~$34$ms for \mdp); this is
unsurprising given our CPU-based implementation of an
expensive transform but nevertheless illustrates \mdp's ability to
process video streams.

\section{Related Work}
\label{sec:relatedwork}

In this section, we discuss related techniques and systems.

\minihead{Streaming and Specialized Analytics} \macrobase is a data analysis system specialized for prioritizing attention in fast data streams. In its architecture, \macrobase builds upon a long history of systems for streaming data and specialized, advanced analytics tasks. A range of systems from both academia~\cite{borealis,telegraphcq} and industry (e.g., Storm, StreamBase, IBM Oracle Streams) provide infrastructure for executing streaming queries. \macrobase adopts dataflow as its execution substrate, but its goal is to provide a set of high-level analytic monitoring operators; in \macrobase, dataflow is a means to an end rather than an end in itself. In designing a specialized engine, we were inspired by several past projects, including Gigascope (specialized for network monitoring)~\cite{gigascope}, WaveScope (specialized for signal processing) ~\cite{wavescope}, MCDB (specialized for Monte Carlo-based operators)~\cite{mcdb}, and Bismarck (providing extensible aggregation for gradient-based optimization)~\cite{bismarck}. In addition, a range of commercially-available analytics packages provide advanced analytics functionality---but, to the best of our knowledge, not the streaming explanation operations we seek here. \macrobase continues this tradition by providing a specialized set of operators for classification and explanation of fast data, which in turn allows new optimizations. We further discuss this design philosophy in~\cite{mb-cidr}.

\minihead{Classification} Classification and outlier detection have an
extensive history; the literature contains thousands of techniques
from communities including statistics, machine learning, data mining,
and information theory~\cite{anomaly-survey1, anomaly-survey2,
  aggarwal-book}. Outlier detection techniques have seen major success
in several domains including network intrusion
detection~\cite{nids1,nids2}, fraud detection (leveraging a variety of
classifiers and techniques)~\cite{fraud1, fraud2}, and industrial
automation and predictive
maintenance~\cite{industrial1,industrial2}. A considerable subset of
these techniques operates over data
streams~\cite{streaming-anomaly1,streaming-anomaly2,streaming-anomaly3,aggarwal-streams}.

At stream volumes in the hundreds of thousands or more events per second, statistical outlier detection techniques will (by nature) produce a large stream of outlying data points. As a result, while outlier detection forms a core component of a fast data analytics engine, it must be coupled with streaming explanation. In the design of \macrobase, we treat the array of classification techniques as inspiration for a modular architecture. In \macrobase's default pipeline, we leverage detectors based on robust statistics~\cite{robust-huber,robust-maronna}, adapted to the streaming context. However, in this paper, we also demonstrate compatibility with detectors from Elki~\cite{elki}, Weka~\cite{weka}, RapidMiner~\cite{rapidminer}, and OpenGamma~\cite{opengamma}.

\minihead{Data explanation} Data explanation techniques assist in summarizing differences between datasets. The literature contains several recent explanation techniques leveraging decision-tree~\cite{failure-trees} and Apriori-like~\cite{patternmining,scorpion} pruning, grid search~\cite{perfaugur,anomaly-localization}, data cubing~\cite{suciu-cube}, Bayesian statistics~\cite{dataxray}, visualization~\cite{vsoutlier,nair-learning}, causal reasoning~\cite{dbsherlock}, and several others~\cite{interpretable, meliou-tutorial, fnt-cleaning,hellerstein-survey}. While we are inspired by these results, none of these techniques executes over streaming data or at the scale we seek. Several exhibit runtime exponential in the number of attributes (which can number in the hundreds of thousands to millions in the fast data we examine)~\cite{dataxray,suciu-cube} and, when reported, runtimes in the batch setting often vary from hundreds to approximately 10K points per second~\cite{scorpion, perfaugur, dataxray} (we also directly compare throughput with several techniques~\cite{dataxray,suciu-cube,failure-trees,scorpion} in Appendix~\ref{apdx:summarycompare}).




To address the demands of streaming operation and to scale to millions of events per second, \macrobase's explanation techniques draw on sketching and streaming data structures (specifically~\cite{cormode-survey,spacesaving,cormode-decay,achao,efraimidis-sampling,itemset-streams,cps-tree,florin-thesis}), adapted to the fast data setting. We view existing explanation techniques as a useful second step in analysis following the explanations generated by \macrobase, and we see promise in adapting these existing techniques to streaming execution at high volume. Given our goal of providing a generic architecture for analytic monitoring, future improvements in streaming explanation should be complementary to our results here.



\section{Conclusions and Future Work}
\label{sec:conclusion}

We have presented \macrobase, a new analytics engine designed to
prioritize attention in fast data streams. \macrobase provides a
flexible architecture that combines streaming classification and data
explanation techniques to deliver interpretable summaries of important
 behavior in fast data streams. \macrobase's default
analytics operators, which include new sampling and sketching
procedures, take advantage of this combination of detection and
explanation and are specifically optimized for high-volume,
time-sensitive, and heterogeneous data streams, resulting in improved
performance and result quality. This emphasis on flexibility,
accuracy, and speed has proven useful in several production
deployments, where \macrobase has already identified previously
unknown behaviors.

\macrobase is available as open source and is under active
development. The system serves as the vehicle for a number of ongoing
research efforts, including techniques for temporally-aware
explanation, heterogeneous sensor data fusion, online non-parametric
density estimation, and contextual outlier detection. Ongoing
production use cases continue to stimulate the development of new
functionality to expand the set of supported domains and leverage the
flexibility provided by \macrobase's pipeline architecture.

\section*{Acknowledgments}
We thank the many members of the Stanford InfoLab, our collaborators
at MIT and Waterloo, Ali Ghodsi, Joe Hellerstein, Mark Phillips, Leif
Walsh, and the early adopters of the MacroBase prototype for providing
feedback on and inspiration for this work. This research was supported
in part by Toyota Research Institute, Intel, the Army High Performance
Computing Research Center, RWE AG, Visa, Keysight Technologies,
Facebook, VMWare, and Philips Lighting, and by the NSF Graduate
Research Fellowship under grants DGE-114747 and DGE-1656518. As
MacroBase is open source and publicly available, there is no
correspondence---either direct or implied---between the use cases
described in this work and the above institutions that supported this
research.

\scriptsize
\selectfont

\bibliography{macrobase} \bibliographystyle{abbrv}

\fontsize{8.45pt}{9.75pt}\selectfont
\selectfont

\section*{APPENDIX}
\setcounter{section}{0}
 \setcounter{subsection}{0}
 \def\thesection{\Alph{section}}

\section{Classification}
\label{appendix:detect}

\minihead{MCD} Computing the exact MCD requires examining all subsets
of points to find the subset whose covariance matrix exhibits the
minimum determinant. This is computationally intractable for even
modestly-sized datasets. Instead, \macrobase adopts an iterative
approximation called FastMCD~\cite{fastmcd}. In FastMCD, an initial
subset of points $S_0$ is chosen from the input set of points
$P$. FastMCD computes the covariance $C_0$ and mean $\mu_0$ of $S_0$,
then performs a ``C-step'' by finding the set $S_1$ of points in $P$
that have the $|S_1|$ closest Mahalanobis distances (to $C_0$ and
$\mu_0$). FastMCD subsequently repeats C-steps (i.e., computes the
covariance $C_1$ and mean $\mu_1$ of $S_1$, selects a new subset $S_2$
of points in $P$, and repeats) until the change in the determinant of
the sample covariance converges (i.e., $det(S_{i-1})-det(S_i)<
\epsilon$, for small $\epsilon$). To determine which dimensions are
most anomalous in MCD, \macrobase uses the corr-max
transformation~\cite{mcd-contributions}.

\minihead{Handling variable ADR arrival
  rates}\label{appendix:variable-fdr} We consider two policies for
collecting samples using an ADR over real-time periods with variable
tuple arrival rates. The first is to compute a uniform sample per
decay period, with decay across periods. This can be achieved by
maintaining an ADR for the stream contents from all prior periods and
a regular, uniform reservoir sample for the current period. At the end
of the period, the period sample can be inserted into the ADR. The
second policy is to compute a uniform sample over time, with decay
according to time. In this setting, given a sampling period (e.g.,
1s), for each period, insert the average of all points.

\minihead{Contamination plot details}\label{apdx:contamination} In
Figure~\ref{fig:discriminatory}, we examine a dataset of
10M points drawn from two distributions: a uniform {\it
  inlier} distribution, with radius 50 centered at the origin,
and a uniform {\it outlier} distribution, with radius 50
centered at (1000, 1000). We varied the proportion of points in each to evaluate the effect of contamination on
the Z-Score, MAD, and MCD (using univariate points for Z-Score and
MAD).

\section{Explanation}

\minihead{Streaming combinations: CPS-tree adaptation} Given the set of recently frequent items, \mdp monitors the attribute stream for frequent attribute combinations by maintaining a frequency-descending prefix tree of attribute values: the CPS-tree data structure~\cite{cps-tree}, with several modifications, which we call the M-CPS-tree. Like the CPS-tree, the M-CPS-tree maintains both the basic FP-tree data structures as well as a set of leaf nodes in the tree. However, in an exponentially damped model, the CPS-tree stores at least one node for every item ever observed in the stream. This is infeasible at scale. As a compromise, the M-CPS-tree only stores items that were frequent in the previous window: at each window boundary, \macrobase updates the frequent item counts in the M-CPS-tree based on its AMC sketch. Any items that were frequent in the previous window but were not frequent in this window are removed from the tree. \macrobase then decays all frequency counts in the M-CPS-tree nodes and re-sorts the M-CPS-tree in frequency descending order (as in the CPS-tree, by traversing each path from leaf to root and re-inserting as needed). Subsequently, attribute insertion can continue as in the FP-tree.

\minihead{Confidence}\label{appendix:confidence} To provide confidence
intervals on its output explanations and prevent false discoveries
(type I errors, our focus here), \mdp leverages existing results from
the epidemiology literature, applied to the \mdp data structures. For
a given attribute combination appearing $a_o$ times in the outliers
and $a_i$ times in the inliers, with a risk ratio of $o$, $b_o$ other
outlier points, and $b_i$ other inlier points, we can compute a
$1-p\%$ confidence interval as:
$$o~\pm~\exp\left ( z_p\sqrt{\frac{1}{a_o}-\frac{1}{a_o+a_i}+\frac{1}{b_o}-\frac{1}{b_o+b_i}}
\right )$$ where $z_p$ is the z-score corresponding to the
$1-\frac{p}{2}$ percentile~\cite{risk-confidence}. For example, an
attribute combination with risk ratio of $5$ that appears in $1\%$ of
$10$M points has a $95$th percentile confidence interval of $(3.93,
6.07)$ ($99$th: $(3.91, 6.09)$). Given a risk ratio threshold of $3$,
\macrobase can return this explanation with confidence.

However, because \mdp performs a repeated set of statistical tests to
find attribute combinations with sufficient risk ratio, \mdp subject to
the multiple testing problem: large numbers of statistical tests are
statistically likely to contain false positives. To address this
problem, \mdp can apply a correction to its intervals. For example,
under the Bonferroni correction~\cite{bonferroni}, if a user seeks a
confidence of $1-p$ and \mdp tests $k$ attribute combinations, \mdp
should instead assess the confidence for $z_p$ at $1-\frac{p}{k}$. We can
compute $k$ at explanation time by recording the number of support
computations.

$k$ is likely to be large as, in the limit, \mdp may examine the power
set of all attribute values in the outliers. However, with fast data,
this is less problematic. First, the pruning power of \mdp's
explanation routine eliminates many tests, thus reducing type I
errors. Second, empirically, many of \macrobase's explanations have
very high risk ratio---often in the tens or hundreds. This is because
many problematic behaviors are highly systemic, meaning large
intervals may still be above the user-specified risk ratio
threshold. Third, and perhaps most importantly, \macrobase analyzes
large streams. In the above example, even with $k=10M$, the $95$th
percentile confidence interval is still $(3.80, 6.20)$. Compared to
medical studies with study sizes in the range of hundreds of samples,
the large volume of data mitigates many of the problems associated
with multiple testing. For example, the same $k=10M$ yields a $95$th
percentile confidence interval of $(0, 106M)$ when applied to a
dataset of only 1000 points, which is effectively meaningless. (This
trend also applies to alternative corrective methods such as the
Benjamini-Hochberg procedure~\cite{fdr}.) Thus, while the volumes of
fast data streams pose significant computational challenges, they can
actually improve the statistical quality of analytics results.


\section{Implementation}
\label{sec:implementation}

In this section, we describe the \macrobase prototype implementation and runtime. As of February 2017, \macrobase's core comprises approximately 9,400 lines of Java, over 7,000 of which are devoted to operator implementation, along with an additional 1,000 lines of Javascript and HTML for the front-end and 7,600 lines of Java for diagnostics and prototype pipelines.

We chose Java due to its high productivity, support for higher-order
functions, and popularity in open source. However, there is
considerable performance overhead associated with the Java virtual
machine (JVM). Despite interest in bytecode generation from high-level
languages such as Scala and .NET~\cite{trill,klonatos2014building}, we
are unaware of any generally-available, production-strength operator
generation tools for the JVM. As a result, \macrobase leaves
performance on the table in exchange for programmer productivity. To
understand the performance gap, we rewrote a simplified \mdp pipeline
in hand-optimized C++. As Table~\ref{table:cpp} shows, we measure an
average throughput gap of 12.76$\times$ for simple queries. JVM code
generation will reduce this gap.

\begin{table}
\center
\scriptsize
\setlength\tabcolsep{3 pt}
\begin{tabular}{c|c|c|c|c|c|c|c}
\cline{2-7}
& LS & TS & ES & AS & FS & MS \\\hline
\multicolumn{1}{|l|}{Throughput (points/sec)} & 7.86M & 8.70M & 9.35M & 12.31M & 7.05M &
6.22M \\
\multicolumn{1}{|l|}{Speedup over Java} & 7.46$\times$ & 24.11$\times$ & 5.24$\times$ &
16.87$\times$ & 5.32$\times$ & 17.54$\times$ \\\hline
\end{tabular}
\caption{Speedups of hand-optimized C++ over Java \macrobase prototype
  for simple queries (queries from Section~\ref{sec:evaluation}).}.
\label{table:cpp}
\end{table}

\macrobase executes operator pipelines via a custom single-core
dataflow execution engine. \macrobase's streaming dataflow decouples
producers and consumers: each operator writes (i.e,. pushes) to an
output stream but consumes tuples as they are pushed to the operator
by the runtime (i.e., implements a
\texttt{consume(OrderedList<Point>)} interface). This facilitates a
range scheduling policies: operator execution can proceed
sequentially, or by passing batches of tuples between
operators. \macrobase supports several styles of pipeline
construction, including a fluent, chained operator API. By default,
\macrobase amortizes calls to \texttt{consume} across several thousand
points, reducing function call overhead. This API also allows stream
multiplexing and is compatible with a variety of existing dataflow
execution engines, including Storm, Heron, and Amazon Streams, which
could act as future execution substrates. We demonstrate
interoperability with several existing data mining frameworks in
Appendix~\ref{apdx:nonfree}.

The \macrobase prototype does not currently implement fault tolerance,
although classic techniques such as state-based checkpointing are
applicable here~\cite{stream-recovery}, especially as \mdp's operators
contain modest state. The \macrobase prototype is also oriented
towards single-core deployment. For parallelism, \macrobase currently
runs one query per core (e.g., one query pipeline per application
cluster in a datacenter). We report on preliminary multi-core
scale-out results in Appendix~\ref{apdx:scaleout}.

The \macrobase prototype and all code evaluated in this paper are available online under a permissive open source license.

\section{Experimental Results}
\label{apdx:expts}

\minihead{Dataset descriptions} \label{appendix:datasets} \cmtd
contains user drives at \cmt, including anonymized metadata such as
phone model, drive length, and battery drain; \telecom contains
aggregate internet, SMS, and telephone activity for a Milanese
telecom; \accidents contains statistics about United Kingdom road
accidents between 2012 and 2014, including road conditions, accident
severity, and number of fatalities; \campaigns contains all US
Presidential campaign expenditures in election years between 2008 and
2016, including contributor name, occupation, and amount; \disburse
contains all US House and Senate candidate disbursements in election
years from 2010 through 2016, including candidate name, amount, and
recipient name; and \liquor contains sales at liquor stores across the
state of Iowa. All but \cmt are $i.)$ publicly accessible, allowing
reproducibility, and $ii.)$ representative of many challenges we have
encountered in analyzing production data beyond \cmt in both scale and
behaviors. While none of these datasets contain ground-truth
labels, we have verified several of the explanations from our queries
over \cmt.

\minihead{Score distribution}\label{apdx:scores} We plot the CDF of
scores in each of our real-world dataset queries in
Figure~\ref{fig:score-cdf}. While many points have high outlier
scores, the tail of the distribution (at the 99th percentile) is
extreme: a very small proportion of points have outlier scores over
over 150. Thus, by focusing on this small upper percentile, \mdp
highlights the most extreme behaviors.

\begin{figure}
\includegraphics[width=.48\columnwidth]{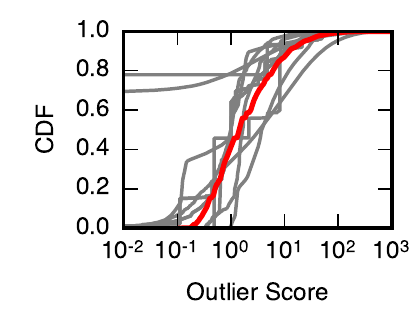} \hfill
\includegraphics[width=.48\columnwidth]{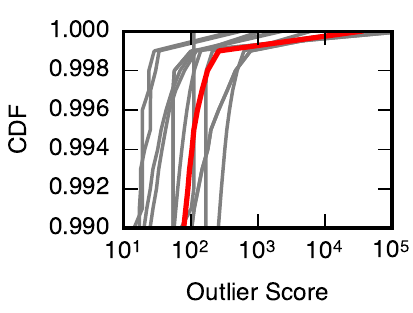}\vspace{-1em}
\caption{CDF of outlier scores for all datasets, with average in red;
  the datasets exhibit a long tail with extreme outlier scores at the
  99th percentile and higher.}
\label{fig:score-cdf}
\end{figure}

\minihead{Varying support and risk ratio}\label{apdx:support-oi} To
understand the effect of support and risk ratio threshold on
explanation, we varied each and measured the resulting runtime and the
number of summaries produced on the EC and MC datasets, which we plot
in Figure~\ref{fig:support-compare}. Each dataset has few attributes
with outlier support greater than 10\%, but each had over 1700 with
support greater than 0.001\%. Modifying the support threshold beyond
0.01\% had limited impact on runtime; most time in explanation is
spent in simply iterating over the inliers rather than maintaining
tree structures. This effect is further visible when varying the risk
ratio, which has less than 40\% impact on runtime yet leads to an
order of magnitude change in number of summaries. Our
default setting of support and risk ratio yields a sensible
trade-off between number of summaries produced and runtime.

\begin{figure}
\includegraphics[width=.48\columnwidth]{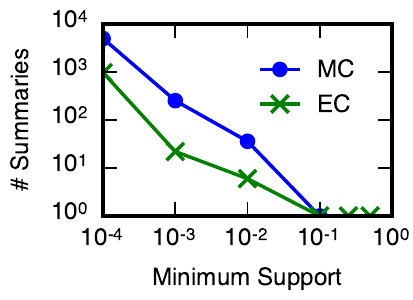}
\hfill \includegraphics[width=.48\columnwidth]{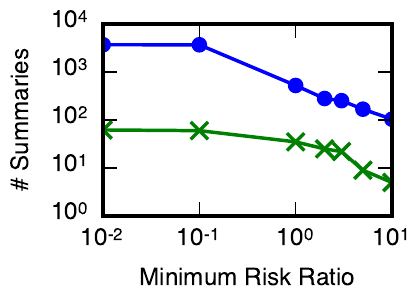}\\
\includegraphics[width=.48\columnwidth]{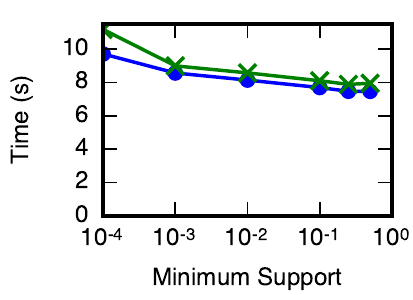} \hfill \includegraphics[width=.48\columnwidth]{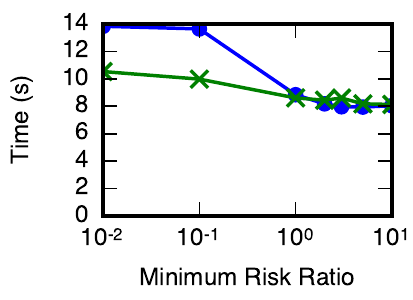}\vspace{-1em}
\caption{Number of summaries produced and summarization time under
  varying support (percentage) and risk ratio.}
\label{fig:support-compare}
\end{figure}

\minihead{Operating on samples}\label{apdx:samples} \mdp periodically
trains models using samples from the input distribution. The
statistics literature offers confidence intervals on the
MAD~\cite{mad-confidence} and the Mahalanobis
distance~\cite{mcd-confidence} (e.g., {for} a sample of size $n$, the
confidence interval of MAD shrinks with $n^{1/2}$), while MCD
converges at a rate of $n^{-1/2}$~\cite{mcd-convergence}. To
empirically evaluate these effects, we measured the accuracy and
efficiency of training models on samples from a $10M$ point
dataset. In Figure~\ref{fig:sample-static}, we plot the outlier
classification accuracy versus sample size for the \cmt queries. MAD
precision and recall are largely unaffected by sampling, allowing a
two order-of-magnitude speedup without loss in accuracy. In contrast,
MCD accuracy is slightly more sensitive due to variance in the sample
selection. This variance is partially offset by the fact that models
are retrained regularly under streaming execution, and the resulting
speedups in both models are substantial.

\minihead{Metric scalability} As Figure~\ref{fig:metric-scalability}
demonstrates, MCD train and score throughput (here, over Gaussian
data) is linearly affected by data dimensionality, encouraging the use
of dimensionality reduction techniques for complex
data.

\begin{figure}
\includegraphics[width=.48\columnwidth]{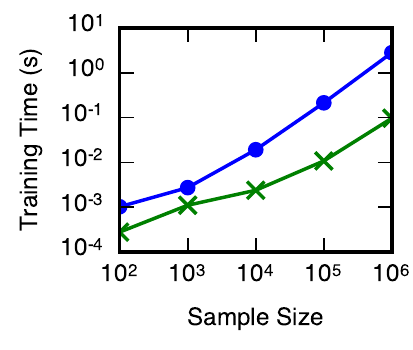}\hfill
\includegraphics[width=.48\columnwidth]{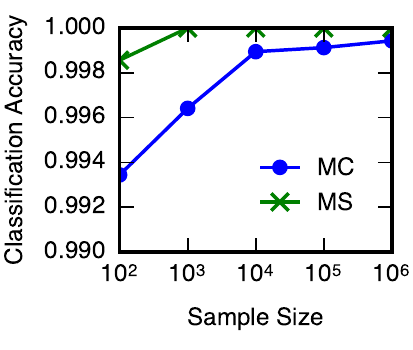}\vspace{-1em}
\caption{Behavior of MAD (MS) and MCD (MC) on samples.}\vspace{-.5em}
\label{fig:sample-static}
\end{figure}

\begin{figure}
\includegraphics[width=\columnwidth]{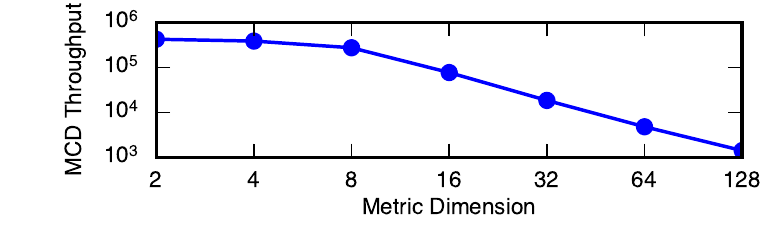}\vspace{-1em}
\caption{MCD throughput versus metric size.}
\label{fig:metric-scalability}
\end{figure}

\minihead{M-CPS and CPS behavior}\label{apdx:mcps} We also investigated the behavior
of the M-CPS-tree compared to the generic CPS-tree. The two data
structures have different behaviors and semantics: the M-CPS-tree
captures only itemsets that are frequent for at least two windows by
leveraging an AMC sketch. In contrast, CPS-tree captures all frequent
combinations of attributes but must insert each point's attributes
into the tree (whether supported or not) and, in the limit, stores
(and re-sorts) all items ever observed in the stream. As a result,
across all queries except ES and EC, the CPS-tree was on average 130x
slower than the M-CPS-tree (std dev: 213x); on ES and EC, the CPS-tree
was over 1000x slower. The exact speedup was influenced by the number
of distinct attribute values in the dataset: \accidents had few
values, incurring 1.3x and 1.7x slowdowns, while \campaigns had many,
incurring substantially greater slowdowns

\minihead{Preliminary scale-out}\label{apdx:scaleout} As a preliminary
assessment of \macrobase's potential for scale-out, we examined \mdp
behavior under a na\"{i}ve, shared-nothing parallel execution
strategy. We partitioned the data across a variable number of cores of
a server containing four Intel Xeon E7-4830 2.13 GHz CPUs and
processed each partition in parallel; upon completion, we return the union of each
core's explanation. As Figure~\ref{fig:scaleout} shows, this strategy
delivers excellent linear scalability. However, as each core
processes a sample of the overall dataset,
accuracy suffers due to both model drift (as in
Figure~\ref{fig:sample-static}) and lack of cross-partition
cooperation in summarization. For example, with 32 partitions spanning
32 cores, FS achieves throughput nearing 29M points per second, with
perfect recall, but only 12\% accuracy. Improving accuracy while
maintaining scalability is the subject of ongoing work.

\begin{figure}
\includegraphics[width=.48\columnwidth]{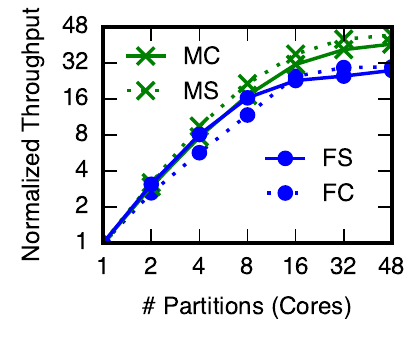}\hfill
\includegraphics[width=.48\columnwidth]{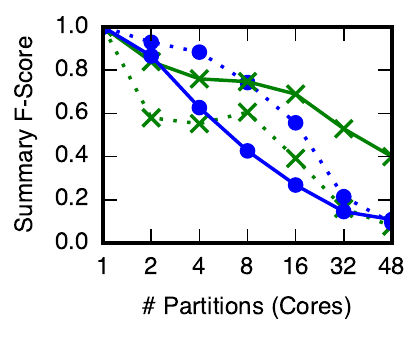}\vspace{-1em}
\caption{Behavior of na\"{i}ve, shared-nothing scale-out.}
\label{fig:scaleout}
\end{figure}

\label{appendix:dbsherlock}

\begin{table}[h!]
\scriptsize
\setlength\tabcolsep{3 pt}
\centering
\begin{tabular}{|llllllllll|}
\hline
\multicolumn{10}{|c|}{TPC-C (QS: one \macrobase query per cluster): top-1: 88.8\%, top-3: 88.8\%}     \\\hline
                                & A1    & A2    & A3    & A4   & A5   & A6   & A7   & A8   & A9   \\
Train top-1 correct (of 9)      & 9     & 9     & 9     & 9    & 9    & 8    & 9    & 9    & 8    \\
Holdout top-1 correct (of 2)    & 2     & 2     & 2     & 2    & 2    & 2    & 2    & 2    & 0    \\ \hline
\multicolumn{10}{|c|}{TPC-C (QE: one \macrobase query per anomaly type): top-1: 83.3\%, top-3: 100\%} \\\hline
                                & A1    & A2    & A3    & A4   & A5   & A6   & A7   & A8   & A9   \\
Train top-1 correct (of 9)      & 9     & 9     & 9     & 9    & 9    & 8    & 9    & 9    & 7    \\
Holdout top-1 correct (of 2)    & 2     & 2     & 2     & 2    & 2    & 1    & 2    & 2    & 0    \\ \hline
\multicolumn{10}{|c|}{TPC-E (QS: one \macrobase query per cluster): top-1: 83.3\%, top-3: 88.8\%}     \\\hline
                                & A1    & A2    & A3    & A4   & A5   & A6   & A7   & A8   & A9   \\
Train top-1 correct (of 9)      & 9     & 9     & 9     & 9    & 9    & 8    & 9    & 9    & 0    \\
Holdout top-1 correct (of 2)    & 2     & 2     & 2     & 2    & 2    & 1    & 2    & 2    & 0    \\ \hline
\multicolumn{10}{|c|}{TPC-E (QE: one \macrobase query per anomaly type): top-1: 94.4\%, top-3: 100\%} \\\hline
                                & A1    & A2    & A3    & A4   & A5   & A6   & A7   & A8   & A9   \\
Train top-1 correct (of 9)      & 9     & 9     & 9     & 9    & 9    & 8    & 9    & 9    & 6    \\
Holdout top-1 correct (of 2)    & 2     & 2     & 2     & 2    & 2    & 1    & 2    & 2    & 2   \\\hline
\end{tabular}
\caption{\mdp accuracy on DBSherlock workload. A1:
  workload spike, A2: I/O stress, A3: DB backup, A4: table restore,
  A5: CPU stress, A6: flush log/table; A7: network congestion; A8:
  lock contention; A9: poorly written query. ``Poor physical design''
  (from~\cite{dbsherlock}) is excluded as the labeled
  anomalous regions did not exhibit significant correlations with any metrics.}
\label{table:dbsherlock}
\end{table}

\minihead{Explanation runtime comparison}\label{apdx:summarycompare}
Following the large number of recent data explanation techniques
(Section~\ref{sec:relatedwork}), we implemented several additional
methods. The results of these methods are not comparable, and prior
work has not evaluated these techniques with respect to one another in
terms of semantics or performance. We do not attempt a full comparison
based on semantics but do perform a comparison based on running time,
which we depict in Table~\ref{table:explain-compare}.  We compared to
a data cubing strategy suggested by Roy and Suciu~\cite{suciu-cube},
which generates counts for all possible combinations (21x slower),
Apriori itemset mining~\cite{patternmining} (over 43x slower), and
Data X-Ray~\cite{dataxray}. Cubing works better for data with fewer
attributes, while Data X-Ray is optimized for hierarchical data; we
have verified with the authors of Data-XRay that, for \macrobase's
flat attributes, Data X-Ray will consider all combinations unless
stopping criteria are met. \macrobase's cardinality-aware explanation
completes fastest for all queries.

\begin{table}
\center
\small
\begin{tabular}{|c|r|r|r|r|r|r|r|r|}
\hline
Query & \multicolumn{1}{c|}{{MB}} & \multicolumn{1}{c|}{{FP}} & \multicolumn{1}{c|}{{Cube}} & \multicolumn{1}{c|}{{DT10}} &
\multicolumn{1}{c|}{{DT100}} & \multicolumn{1}{c|}{{AP}} &
\multicolumn{1}{c|}{{XR}}\\\hline
LC &  {1.01}  &  4.64  & DNF &  7.21  &  77.00  & DNF & DNF \\
TC &  {0.52}  &  1.38  &  4.99  &  10.70  &  100.33  &  135.36  & DNF \\
EC &  {0.95}  &  2.82  &  16.63  &  16.19  &  145.75  &  50.08  & DNF \\
AC &  {0.22}  &  0.61  &  1.10  &  1.22  &  1.39  &  9.31  &  6.28  \\
FC &  {1.40}  &  3.96  &  71.82  &  15.11  &  126.31  &  76.54  & DNF \\
MC &  {1.11}  &  3.23  & DNF &  11.45  &  94.76  & DNF & DNF \\\hline
\end{tabular}
\caption{Running time of explanation algorithms (s) for
  each complex query. MB: \macrobase, FP: FPGrowth, Cube: Data
  cubing; DT\textit{X}: decision tree, maximum depth $X$; AP: A-Aprioi;
  XR: Data X-Ray. DNF: did not complete in 20 minutes.}
\label{table:explain-compare}
\end{table}

\minihead{Compatibility with existing frameworks}
\label{apdx:nonfree} We implemented several additional \macrobase
operators to validate interoperability with existing data mining
packages. We were unable to find a single framework that implemented
both unsupervised outlier detection and data explanation and had
difficulty locating streaming implementations. Nevertheless, we
implemented two \macrobase outlier detection operators using Weka
3.8.0's \texttt{KDTree} and Elki 0.7.0's \texttt{SmallMemoryKDTree},
an alternative FastMCD operator based on a recent RapidMiner extension
(\texttt{CMGOSAnomalyDetection})~\cite{rapidminer-mcd}, an alternative
MAD operator from the OpenGamma 2.31.0, and an alternative
FPGrowth-based summarizer based on SPMF version v.0.99i. As none of
these packages allowed streaming operation (e.g., Weka allows adding
points to a \texttt{KDTree} but does not allow removals, while Elki's
\texttt{SmallMemoryKDTree} does not allow modification), we
implemented batch versions. We do not perform accuracy comparisons
here but note that the kNN performance was substantially slower
($>$100x) than \mdp's operators (in line with recent
findings~\cite{rapidminer-mcd}) and, while SPMF's operators were
faster than our generic FPGrowth implementation, SPMF was still
2.8$\times$ slower than \macrobase due to \mdp's cardinality-aware
optimizations. The primary engineering overheads came from adapting to
each framework's data formats; however, with a small number of utility
classes, we were able to easily compose operators from different
frameworks and also from \macrobase, without modification. Should
these frameworks begin to prioritize streaming execution and/or
explanation, this interoperability may prove fruitful in
the future.

\end{document}